\documentclass[twocolumn,numberedappendix]{aastex62}

\usepackage{times}
\usepackage{graphicx}
\usepackage{epstopdf}
\usepackage{subfigure}
\usepackage{color}
\usepackage{verbatim}
\usepackage{bm}% bold math
\usepackage{hyperref}
\usepackage{CJK}

\hypersetup{colorlinks = true, linkcolor = red, anchorcolor = red, citecolor = blue, filecolor = blue, urlcolor = blue}

\shorttitle{Eccentric planet in a dusty disk}
\shortauthors{Li et al.}

\begin{document}
\begin{CJK*}{UTF8}{gbsn}

%\title{Planet-disk Interaction: the Effect of an Eccentric Super-Earth on the Dust Dynamics}
%\title{Orbital Evolution of Eccentric Super-Earths in Dusty Protoplanetary Disks and Observational Implications}

\title{On the Dust Signatures Induced by Eccentric Super-Earths in Protoplanetary Disks}

\correspondingauthor{Ya-Ping Li}
\email{leeyp2009@gmail.com}

\author[0000-0002-7329-9344]{Ya-Ping Li(李亚平)}
\affil{Theoretical Division, Los Alamos National Laboratory, Los Alamos, NM 87545, USA}
\author[0000-0003-3556-6568]{Hui Li(李晖)}
\affil{Theoretical Division, Los Alamos National Laboratory, Los Alamos, NM 87545, USA}
\author[0000-0002-4142-3080]{Shengtai Li(李胜台)}
\affil{Theoretical Division, Los Alamos National Laboratory, Los Alamos, NM 87545, USA}
\author[0000-0001-5466-4628]{Douglas N. C. Lin(林潮)}
\affil{Department of Astronomy and Astrophysics, University of California, Santa Cruz, CA 95064, USA}

\begin{abstract}
{We investigate the impact of a highly eccentric 10 $M_{\rm \oplus}$ (where $M_{\rm \oplus}$ is the Earth mass) planet embedded in a dusty protoplanetary disk on the dust dynamics and its observational implications.  By carrying out high-resolution 2D gas and dust two-fluid hydrodynamical simulations, we find that the planet's orbit can be circularized at large radii. 
After the planet's orbit is circularized, partial gap opening and dust ring formation happen close to  the planet's  circularization radius, which can explain the observed gaps/rings at the outer region of disks. When the disk mass and viscosity become low,  we find that  an  eccentric planet can even open gaps and produce dust rings close to the
pericenter and apocenter radii before its circularization.
This offers alternative scenarios for explaining the observed dust rings and gaps in protoplanetary disks.
A lower disk viscosity is favored to produce brighter rings in
observations.
An eccentric planet can also potentially slow down the dust radial drift in the outer region of the disk
when the disk viscosity is low ($\alpha \lesssim2\times10^{-4}$) and the circularization is
faster than the dust radial drift.}
\end{abstract}

\section{Introduction}

Strong observational evidence  for the existence of multiple dust rings/gaps
in disks has emerged from high-resolution observations with the Atacama Large Millimeter Array (ALMA),
e.g., HL Tau \citep{ALMA2015}, TW Hya \citep{Andrews2016,Tsukagoshi2016}, HD 163296 \citep{Isella2016},
AA Tau \citep{Loomis2017}, Elias 24 \citep{Cieza2017,Cox2017,Dipierro2018}, AS 209 \citep{Fedele2018},
GY 91 \citep{Sheehan2018}, V1094 Sco \citep{Ansdell2018,Terwisga2018}, MWC 758 \citep{Boehler2018,Dong2018a},
including
several recent surveys of young disks \citep{Long2018,Andrews2018,vanderMarel2019}.
A large fraction of these rings are located at the outer region of the disks, i.e., $\sim50-100\ {\rm au}$.
Planet-disk interaction (e.g., \citealt{Dipierro2015,Dong2015,Pinilla2015,Jin2016,Dong2017,Dong2018b}) is one
of the main mechanisms that are proposed to produce these rings, although several other scenarios
have also been proposed (e.g., \citealt{Pinilla2012,Takahashi2014,Gonzalez2015,Suriano2017,Suriano2018,Zhang2015,Okuzumi2016,Miranda2017}).
It is generally thought that forming (sometimes massive) planets \textit{in situ} at
such large disk radii  is quite difficult.
It is then worthwhile to explore whether there are mechanisms
that deliver planets to such large radii even though other may be born at relatively small disk radii.

Observationally, many exoplanets discovered so far are found to be in highly
eccentric orbits\footnote{http://exoplanets.org/}.
Planet-disk and/or planetary scattering may be responsible for the eccentricity for
the inner ``hot Jupiter" such as CI Tau \citep{Duffell2015,Rosotti2017} while delivering
other planets into the outer region of the disk by the eccentricity pumping,
where three rings were discovered \citep{Clarke2018}.
Multiple planet-planet scattering is a promising mechanism that delivers planets from
a few au where they were likely born to the outer disk region by exciting the planet's orbit to a high eccentricity,
which then circularizes at a large radius.
In addition,
many previous studies have examined the eccentricity excitation of massive gas giants \citep[e.g.,][]{Rasio1996,Juric2008,Marzari2010,Moeckel2012,Lega2013,Rosotti2017}.
Furthermore, an eccentric orbit could also be  the relics due to the planet-disk interaction \citep[e.g.,][]{Papaloizou2001,Goldreich2003,Bitsch2013,Duffell2015}, secular chaos \citep{Wu2011}, Kozai-Lidov oscillations \citep{Kozai1962,Lidov1962,Takeda2005}. In the planet-disk interaction scenario, whether the eccentricity can be excited or damped depends on the dominance of Lindblad and co-rotation resonance torque
\citep{Goldreich2003,Duffell2015}. It is found that a Jupiter mass planet can pump up its orbital eccentricity
due to the interaction with the disk \citep{DAngelo2006}.

In this study, we investigate the impact of a highly eccentric super-Earth on the dust distribution during its orbital evolution.
(The effect of an eccentric massive giant will be studied in a future work.)
Our main goal of this paper is to explore whether an eccentric super-Earth can
leave observational features in a dusty disk, and how these features can be used to infer the planetary properties. Previous studies on eccentric planets mainly focused on the planet migration and circularization timescales, which have either used the linear theory
when the eccentricity is small \citep[e.g.,][]{Goldreich1978,Goldreich1980,Papaloizou2000,Papaloizou2001,Tanaka2004,Duffell2015},
or the non-linear simulations with an initially large eccentricity \citep[e.g.,][]{Cresswell2007,Bitsch2013},
but the influence on dust has not been explored self-consistently.

The paper is organized as follows. The numerical simulation setup will be described in
Section~\ref{sec:method}. We present the results in Section~\ref{sec:results},
and discussions and conclusions are given in Section~\ref{sec:conclusions}.

\section{Method}\label{sec:method}

We use the LA-COMPASS code \citep{Li2005,Li2009,Fu2014} to simulate the coupled
gas-dust and planet dynamics. The main ingredient of our model is that one planet
with a mass of $M_{\rm p}=10\ M_{\oplus}$ in an eccentric orbit,
where $M_{\oplus}$ is the Earth mass, is embedded in the dusty disk.
The planet orbits around the star with an initial orbital eccentricity $e_{\rm p}$
and a semi-major axis of $a_{\rm p}$. The planet is initially located at the pericenter
of the orbit. After 10 orbital evolution with a fixed orbit, the planet is freely released by
considering the dynamical interaction between the disk and the planet. 
For all models listed in Table~\ref{tab:para}, we adopt a softening length of $0.7h_{\rm g}$ for the planet's potential,
where $h_{\rm g}$ is the disk gas scale height.
These runs are used to understand the dependence of orbital evolution of
eccentric planets on disk gas surface density, initial orbital eccentricity, disk viscosity, dust
particle sizes, and their consequent influence on the dust dynamics as well.

A disk, extending from 10 to 500 AU, around a pre-main sequence (PMS) star
with a mass of $M_{\star} = 1.0~M_{\sun}$ is considered.  The disk gas surface density is initially
described by
\begin{equation}\label{eq:gas}
  \Sigma_{\rm g}(r)=\Sigma_{0}\left(\frac{r}{r_{\rm c}}\right)^{-\gamma}\exp\left[-\left(\frac{r}{r_{\rm c}}\right)^{2-\gamma}\right],
\end{equation}
where $r_{\rm c}=150~\rm au$ and $\gamma=1.0$. The normalization of gas surface density $\Sigma_{0}$ is treated as one main free parameter to determine the total disk mass. We have considered two disk masses with $M_{\rm disk}=0.02\ {M_{\odot}}$ and $5\times10^{-3}\ M_{\odot}$, or equivalently $\Sigma_{0}=1.4\ {\rm g\ cm^{-2}}$ and $0.36\ {\rm g\ cm^{-2}}$, respectively.
Disk self-gravity is considered for all models \citep{Li2009} since the minimum Toomre Q parameter can reach $\sim2$ after the planet is circularized.

We choose a locally isothermal equation of state (EoS) with the sound speed $c_{\rm s}$ given by
\begin{equation}\label{eq:cs}
  \frac{c_{\rm s}}{v_{\rm K}}=\frac{h_{\rm g}}{r}=h_{0}\left(\frac{r}{r_{0}}\right)^{0.25},
\end{equation}
where $r_{0}=50$ au. And $h_{0}=0.03$ is adopted as a typical value for the sound speed.
$v_{\rm K}$ is the Keplerian velocity. Note that Equation~(\ref{eq:cs}) also expresses the radial
profile of the gas scale height $h_{\rm g}/r$. The effect for a different adiabatic EoS is discussed
in detail in Appendix~\ref{sec:app_ad}. The disk viscosity is adopted from the
Shakura-Sunyaev prescription $\nu_{\rm g}=\alpha_{\rm vis} c_{\rm s}h_{\rm g}$ with
a constant $\alpha_{\rm vis}$ across the whole disk \citep{Shakura1973}.
A single dust species with its size $s_{\rm d}$ as listed in Table~\ref{tab:para} is included in the simulations.
We assume that the initial dust surface density $\Sigma_{\rm d}$ follows the gas profile
$\Sigma_{\rm g}$, with an initial dust-to-gas ratio of $1\%$. In the Epstein regime for
the disk and dust parameters of interest here, the Stokes number of the particles
with dust radius $s_{\rm d}$ in the mid-plane of the disk is defined as
\begin{equation}\label{eq:st}
  {\rm St} = \frac{\pi\rho_{\rm s}s_{\rm d}}{2\Sigma_{\rm g}},
\end{equation}
where $\rho_{\rm s}$ is the solid density of the dust particle. ${\rm St}(r_{0})=0.01$ for our fiducial model, which is the main parameter that controls the radial drift of dust.

We solve the hydrodynamics equations with a high-resolution 2D grid
of $(n_{r},n_{\phi})=3072\times3072$ in the radial and azimuthal direction. We have tested that an even higher resolution of $(n_{r},n_{\phi})=6144\times6144$ does not change the results significantly. 
A fixed boundary condition is applied to the gas, while an outflow inner
boundary condition and outer boundary condition are imposed on the dust,
which allow the dust flow out/in from the boundary depending on its radial velocity.

In order to identify the observational features from the dust, we follow
\citet{Li2019} to obtain the dust continuum emission at mm wavelengths using the \texttt{RADMC-3D} package \citep{Dullemond2012}.
To convert the 2D dust surface density produced from hydrodynamical simulation into a 3D grid, we adopt a dust scale height $h_{\rm d}=h_{\rm g}\min\left({1,\sqrt{\frac{\alpha_{\rm vis}}{\min(0.5,\rm St)(1+\rm St^2)}}}\right)$ by considering the dust vertical settling \citep{Birnstiel2010a}.
We use $400\times400$ uniform grids in the $r-\phi$ plane and
40 uniform grids in the $\theta$ direction between $70^{\circ}-90^{\circ}$ with
a mirror symmetry in the equatorial plane. This grid can recover all the dust mass from
2D hydrodynamics within a $5\%$ uncertainty. A larger grid number does not change the results.
We then run \texttt{RADMC-3D} simulations to compute the dust temperature $T_{\rm d}(r,\phi, \theta)$
contributed by the stellar radiation.
With the dust temperature $T_{\rm d}$ obtained from \texttt{RADMC-3D} and the dust surface
density $\Sigma_{\rm d}$ for  dust species derived from hydrodynamical simulations, we can calculate the dust continuum by ray-tracing with \texttt{RADMC-3D}.

\begin{table}[htbp]
  \begin{center}
  \caption{\bf Model parameters and circularization time}\label{tab:para}
  \begin{tabular}{lccccc|c}
     \hline\hline
     model & $\Sigma_{0}$ &  $\alpha_{\rm vis}$  & $e_{\rm p}$ & $a_{\rm p}$ & $s_{\rm d}$ & $t_{\rm circ}$\\
                & $(\rm g~cm^{-2})$ &  &    & (au) & (mm) & ({\rm orbits})$^{\dagger}$\\
     \hline
     fiducial & 1.4 & $2\times10^{-4}$ &  0.8 & 100 & 0.2 & 2500\\
     HigVis & 1.4 & $1\times10^{-3}$ &  0.8 & 100 & 0.2 & 2500\\
    % LowE & 1.4 & $2\times10^{-4}$ &  0.5 & 40 & 0.2 & 400\\
     LowMg & 0.36 & $2\times10^{-4}$ &  0.5 & 40 & 0.2 & 1600\\
     S2 & 1.4 & $2\times10^{-4}$ &  0.8 & 100 & 0.05 & 2500\\
     S3 & 1.4 & $2\times10^{-4}$ &  0.8 & 100 & 0.5 & 2500\\
     \hline\hline
   \end{tabular}
   \end{center}
   \tablecomments{
$e_{\rm p}$ and $a_{\rm p}$ are chosen to keep the same pericenter radius of the planet [$a_{\rm p}(1-e_{\rm p})$] for different models here.
$^{\dagger}$ Circularization time measured in unit of orbits at $r_0 = 50$ au. 
}
  \end{table}

\section{Results}\label{sec:results}

\subsection{Planet dynamics}

We first consider one planet with an initial eccentricity of $e_{\rm p}=0.8$ as our fiducial run.

\begin{figure}[htbp]
%\vbox to3.2in{\rule{0pt}{3.2in}} \special{psfile=fig1.EPS voffset=0 hoffset=0 vscale=80 hscale=80 angle=0}
%\centering
\begin{center}
\includegraphics[width=0.5\textwidth]{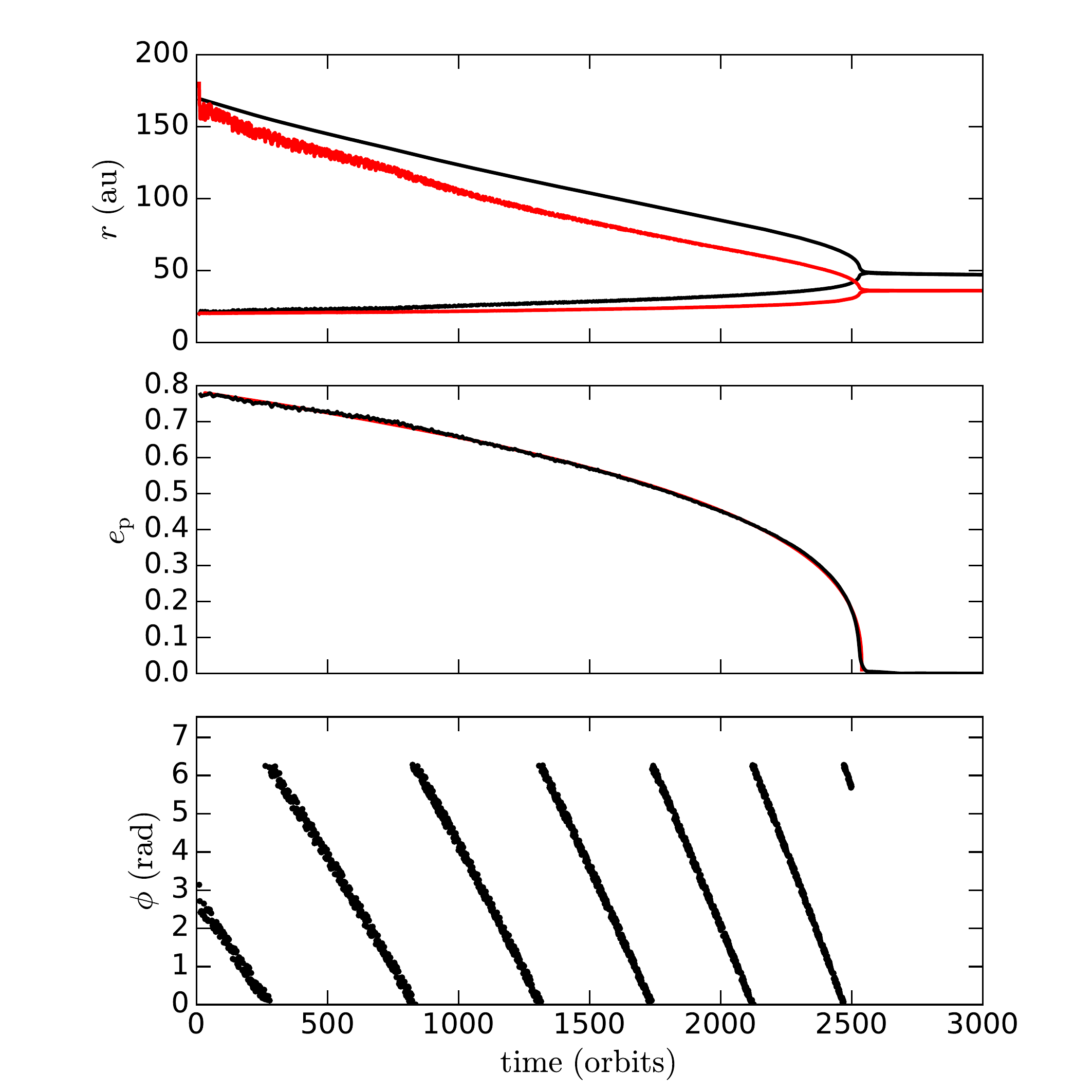}
\end{center}
\caption{Planet orbit circularization for the fiducial model.
Upper panel: black dots show both the planet's pericenter and apocenter locations from the star as a function of time. Red dots show the expected $a_{\rm min}$ and $a_{\rm max}$ if the angular momentum of the planet was conserved. A larger $a_{\rm min}$ and $a_{\rm max}$ compared with the red dots indicate the transfer of angular momentum from the disk to the planet. Middle panel: the evolution of planet's orbital eccentricity as a function of orbital period (black line). The red line shows the fit with $de_{\rm p}/dt= - Ae_{\rm p}^{-\alpha}$, where the best-fitted $\alpha=1.83$. Lower panel: the pericenter phase of the planet orbit as a function of time. The periodical variation of the pericenter's phase suggests a precession period of $\sim450$ orbits. }\label{fig:circ}
\end{figure}

The time evolution of the planet's stellocentric radius (black) is shown in the upper panel of Figure~\ref{fig:circ}.
The specific angular momentum of the planet with an eccentric orbit is defined as
$J_{\rm p} = \sqrt{GM_{\star}a_{\rm p}(1-e_{\rm p}^2)}$. The pericenter and apocenter of the orbit are linked
to the planetary angular momentum $J_{\rm p}$ as $a_{\rm min}=J_{\rm p}^2/[GM_{\star}(1+e_{\rm p})]$
and $a_{\rm max}=J_{\rm p}^2/[GM_{\star}(1-e_{\rm p})]$, respectively.
We also plot the eccentricity $e_{\rm p}$ evolution in the middle panel of Figure~\ref{fig:circ},
which shows that it is damped over $\sim 2500$ orbits.
The increase of the pericenter $a_{\rm min}=a_{\rm p}(1-e_{\rm p})$ and the
decrease of the apocenter $a_{\rm max}=a_{\rm p}(1+e_{\rm p})$ are consistent with the
damping of the eccentricity
with $e_{\rm p}=(a_{\rm max}-a_{\rm min})/(a_{\rm max}+a_{\rm min})$.

\begin{figure*}[htb]
%\vbox to3.2in{\rule{0pt}{3.2in}} \special{psfile=fig1.EPS voffset=0 hoffset=0 vscale=80 hscale=80 angle=0}
%\centering
\begin{center}
\includegraphics[width=0.38\textwidth]{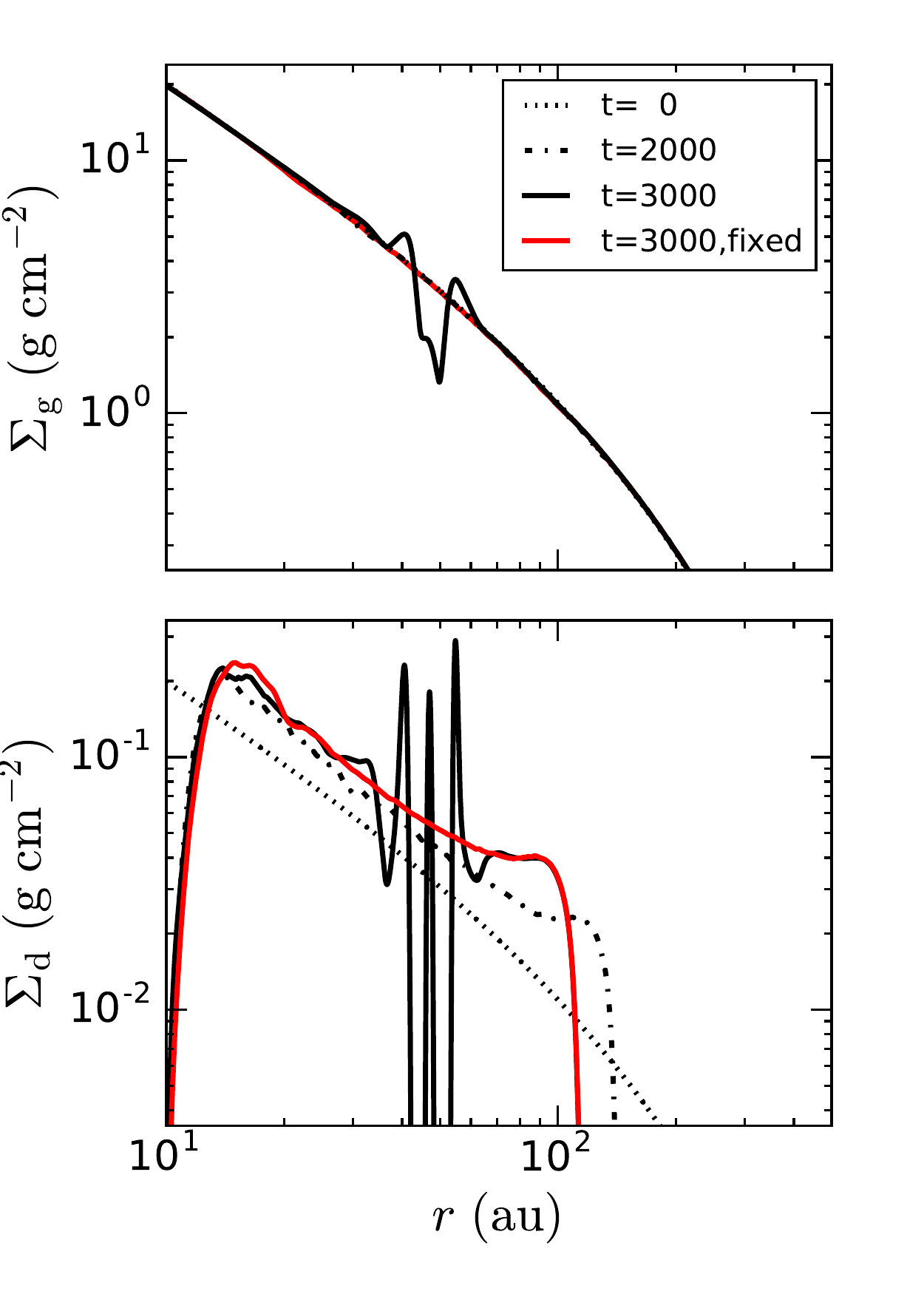}
\includegraphics[width=0.6\textwidth]{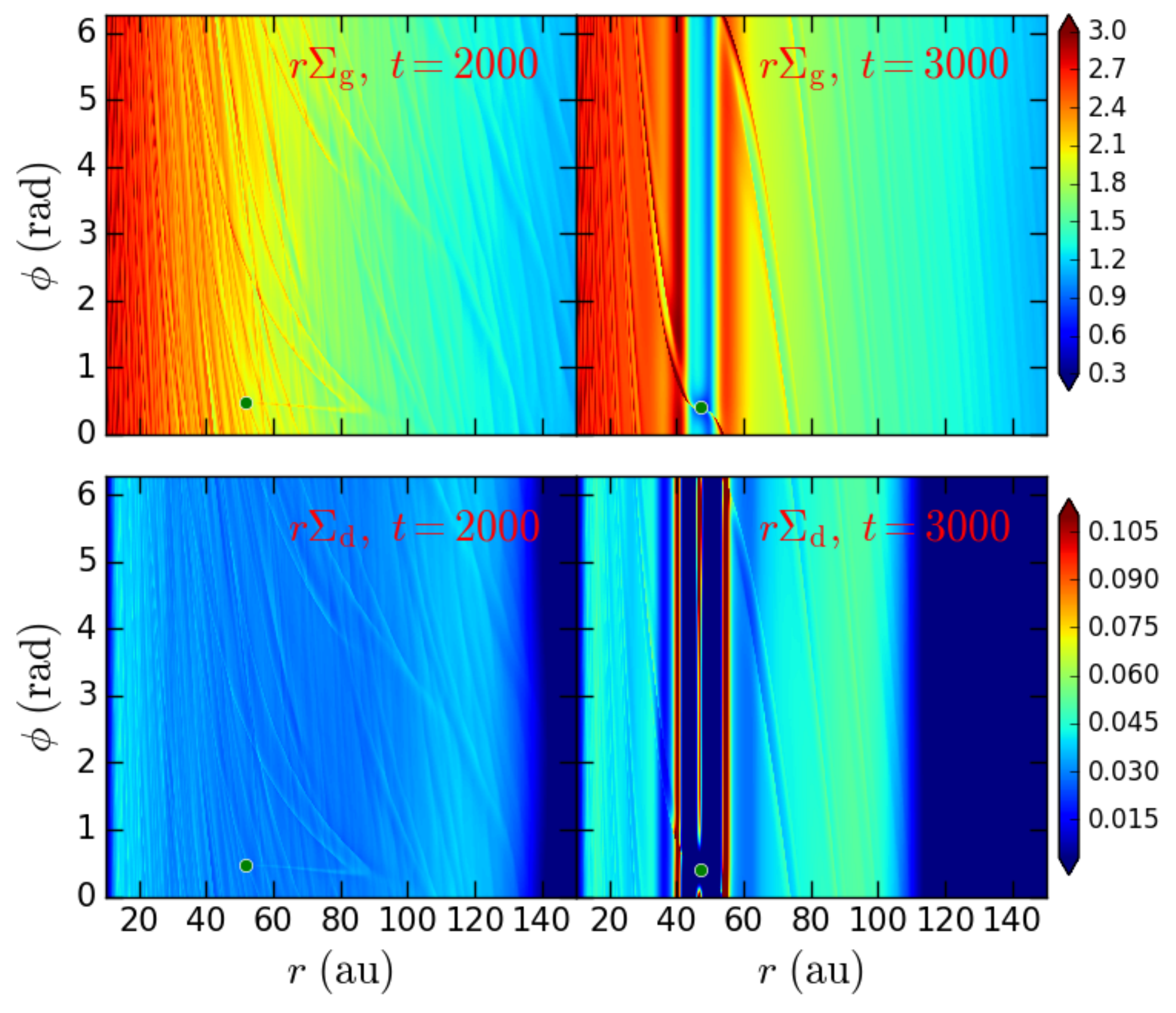}
\end{center}
\caption{Left panels: Azimuthally-averaged gas (upper) and dust (lower) radial profiles for
the fiducial model at different times.  The red lines show the radial profiles of gas and dust at 3000 orbits when the planet is fixed to the initial eccentric orbit. Middle panels: 2D gas and dust surface density
(multiplied by the radius) for the same model at 2000 orbits (before planet circularization).
The planet is shown as the green filled circle. The upper (lower) panel corresponds to gas  (dust).
Right panels: Same as the middle panels but for 3000 orbits (after planet circularization).
Orbital time are all measured at $r_{0}=50$ au, so 3000 orbits correspond
to $\sim1.1$ Myr.}\label{fig:density}
\end{figure*}

We also show the expected $a_{\rm min,l}$ and $a_{\rm max,l}$ (red) in the upper
panel of Figure~\ref{fig:circ} if the planet's orbital angular momentum was conserved
during the whole evolution.
The time derivative of the planet angular momentum is
\begin{equation}\label{eq:Jpdot}
  \frac{\dot{J}_{\rm p}}{J_{\rm p}}=\frac{1}{2}\frac{\dot{a}_{\rm p}}{a_{\rm p}}-\frac{e_{\rm p}^{2}}{1-e_{\rm p}^{2}}\frac{\dot{e}_{\rm p}}{e_{\rm p}}.
\end{equation}
As the planet typically migrates inward with $\dot{a}_{\rm p}<0$ even for a high $e_{\rm p}$, whether the planet gains or loses
orbital angular momentum depends on the value of $e_{\rm p}$.  For large initial $e_{\rm p}$,
the planet gains orbital angular momentum while loses eccentricity.
This is
consistent with the analytic works \citep{Papaloizou2000,Ida2019} and N-body simulations \citep{Cresswell2008}
in the supersonic case (i.e., $e_{\rm p}\gg h_{\rm 0}$).

The gain of planet's orbital angular momentum yields a
circularization radius $R_{\rm circ}$ of the planet at
$\sim50$ au, which is a factor of $\sim1.4$ larger than the value ($36$ au)
if the planet's angular momentum was conserved. This circularization location is also much farther away from the initial pericenter location (20 au).
This presents an interesting case while an initially eccentric planet could eventually be located
at a relatively large stellocentric radius on a circular orbit.

We fit the eccentricity evolution with a power-law form of $de_{\rm p}/dt=-Ae_{\rm p}^{-\alpha}$,
which leads to a timescale $t_{\rm circ}\equiv \int \frac{de_{\rm p}}{de_{\rm p}/dt}=e_{\rm p,0}^{\alpha+1}/A(\alpha+1)$\footnote{This definition of circularization timescale is different from those in previous works by a factor of $1/(\alpha+1)$ \citep[e.g.,][]{Papaloizou2000}.}.
The fitting result is shown as the red line in the middle panel with $\alpha=1.83$ and the corresponding circularization timescale $t_{\rm circ}\simeq2500$ orbits at 50 au, i.e., $0.9$ Myr,
consistent with the circularization timescale of $0.7 - 1.1$ Myr derived from
previous studies \citep{Papaloizou2000, Cresswell2008, Muto2011, Ida2019}, using our model parameters.

Our simulation also suggests a scaling relation of $t_{\rm circ}\propto e_{\rm p}^{2.8}$, consistent with the
previous results of $t_{\rm circ}\propto e_{\rm p}^{3.0}$ in high $e_{\rm p}$ regime
\citep{Papaloizou2000,Cresswell2008,Ida2019}. Meanwhile,
the pericenter of the orbit precesses with a period of 450 orbits, as shown in the lower panel of Figure~\ref{fig:circ}.

\subsection{Gas and dust evolution}

An eccentric planet during and after cirularization could leave imprint on the gas and dust distributions.
The azimuthal-averaged gas and dust surface density radial profiles are shown in Figure~\ref{fig:density}.
We select three different times to show the temporal evolution of the dust and gas distributions.
Initially, both the gas and dust follow the distribution as in Equation~(\ref{eq:gas}).
At 0.7 Myr (2000 orbits), which corresponds to the time before the planet is circularized,
the gas is only marginally disturbed in the whole disk. The dust also has a relatively smooth
distribution except for the truncation at 150 au and a steep drop in the inner boundary.
Such a truncation at the outer edge, which shrinks with time gradually, is due to the dust radial drift, while the decrease of the dust in
the inner edge is due to the open boundary we choose.
This smooth gas/dust distribution is because of the weak dynamical friction
force induced for a higher eccentric planet \citep{Muto2011}.
The pericenter precession of the planet mentioned above makes its interaction with disk even weaker,
which is attributed to a longer time for the planet to encounter the same region of the disk,
leading to the recovering of the gas from the disturbance with the viscous timescale.

After the planet is circularized around 2500 orbits, the disk is strongly perturbed around the
planet orbit, similar to the situation with a planet on a circular orbit. Two rings sandwiching
the gap induced by the planet gradually appear in the gas radial profiles.
Accordingly, the dust can be trapped in the two gas bump regions due to the positive
pressure gradient near the edges of the gas gap.  The density contrast of the gap and
rings becomes gradually higher with time because of the continuous dust trapping in the rings,
shown as the solid line in the left panel of Figure~\ref{fig:density}, which corresponds to the
time at 3000 orbits (i.e., 1.1 Myr). A higher dust-to-gas ratio, $\sim 0.1$, is reached for the dust ring.

The 2D distributions for the gas and dust at two evolution stages (2000 and 3000 orbits)
are shown in the middle and right panels of Figure~\ref{fig:density}. We multiply the gas
and dust surface density by the radial distance $r$ for  illustration purpose.
Before the planet circularization (2000 orbits), both the gas and dust are marginally disturbed
as shown in middle panels of Figure~\ref{fig:density}, although some weak spiral
features exist around the planet.
At 3000 orbits when the planet is in a circular orbit, a prominent gap at the circularization
radius is shown in the gas distribution. A deep gap sandwiched by two bright rings then
shows up at the same location for the dust distribution. The third dust ring also
exists at the planetary co-orbital radius.
The three rings are located at $\sim55\ \rm au$, $\sim 47\ \rm au$ and $\sim41\ \rm au$.
The two rings separated by the planet are close to a pair of 3:2 resonance. 

\begin{figure}[htbp]
%\vbox to3.2in{\rule{0pt}{3.2in}} \special{psfile=fig1.EPS voffset=0 hoffset=0 vscale=80 hscale=80 angle=0}
%\centering
\begin{center}
\includegraphics[width=0.45\textwidth]{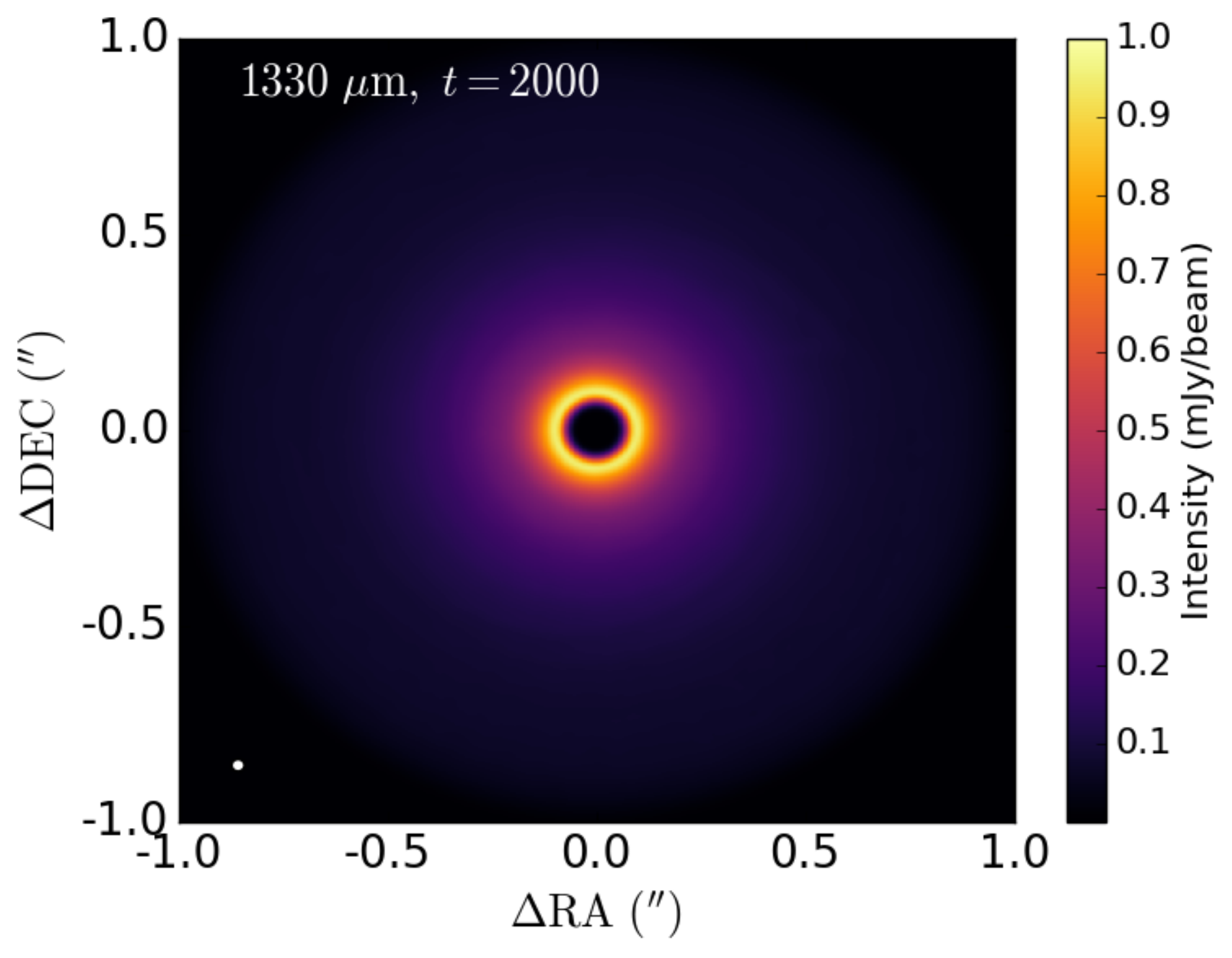}
\includegraphics[width=0.45\textwidth]{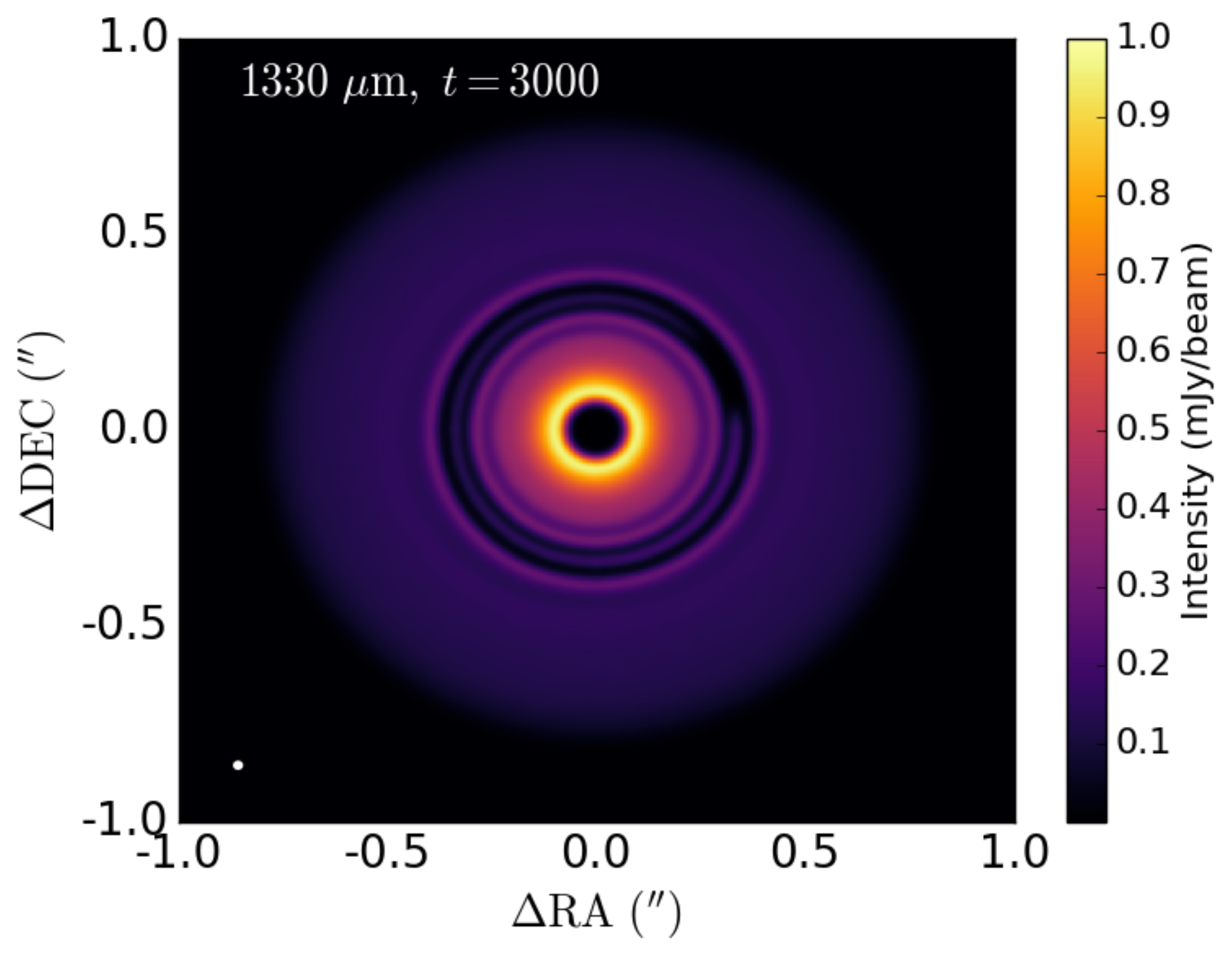}
\end{center}
\caption{Dust continuum image at 1330 $\mu$m for two different evolution stages.  The upper one is prior to the planet circularization (2000 orbits), while the bottom one is after the circularization (3000 orbits). The planet can only have a significant impact on the dust distribution (i.e., gap/rings) after its circularization.}\label{fig:image}
\end{figure}

For the purpose to compare with ALMA observations, we use \texttt{RADMC-3D} package to
produce the 1.33 mm dust continuum, and convolve them with  a gaussian beam of
$0.03^{\prime\prime}\times0.03^{\prime\prime}$.  The disk is assumed to be at a
distance of $140\ {\rm pc}$. This beam size is close to the recent highest resolution
by ALMA \citep{Andrews2018}. Similar to the discussion above, two snapshot at 2000
and 3000 orbits are chosen to produce the  dust image as shown in Figure~\ref{fig:image}.
The featureless image at 2000 orbits is consistent with the smooth dust density distribution.
After the planet orbit circularization, remarkable ringed-structures in the dust density distribution
(a gap plus several rings around the planet circularization radius) can  be recovered.
The dust ring location, which is close to the circularization radius, 
can be in the outer region of the disk because of the angular momentum gaining during the circularization process. This can thus provide a natural explanation for the rings observed at large radii.

We have also tried the case where the planet is kept on a
fixed eccentric orbit (i.e., no planet circularization and precession).
We find that the gas and dust surface density distributions do not show any visible
features and no gaps/rings are produced, shown as the red lines in the left panels of Figure~\ref{fig:density}.
The features shown in Figures~\ref{fig:density} and \ref{fig:image} can only be produced after the planet circularization.

\subsection{Other models}

As we have shown that the dust and gas dynamics are strongly influenced
by the planet circularization process, we explore how the final dust distribution
depends on the planet, disk, and dust parameters, e.g., disk viscosity $\alpha_{\rm vis}$, and disk gas surface density
$\Sigma_{0}$, and dust size $s_{\rm d}$. Models with different parameters are listed in
Table~\ref{tab:para}. For the purpose of completeness, we have also explored the dependence
of the circularization timescale on the disk scale height in high $e_{\rm p}$ and
low $e_{\rm p}$ regimes in  Appendix~\ref{sec:app_tcirc}.

\begin{figure*}[htbp]
%\vbox to3.2in{\rule{0pt}{3.2in}} \special{psfile=fig1.EPS voffset=0 hoffset=0 vscale=80 hscale=80 angle=0}
%\centering
\begin{center}
\includegraphics[width=0.45\textwidth]{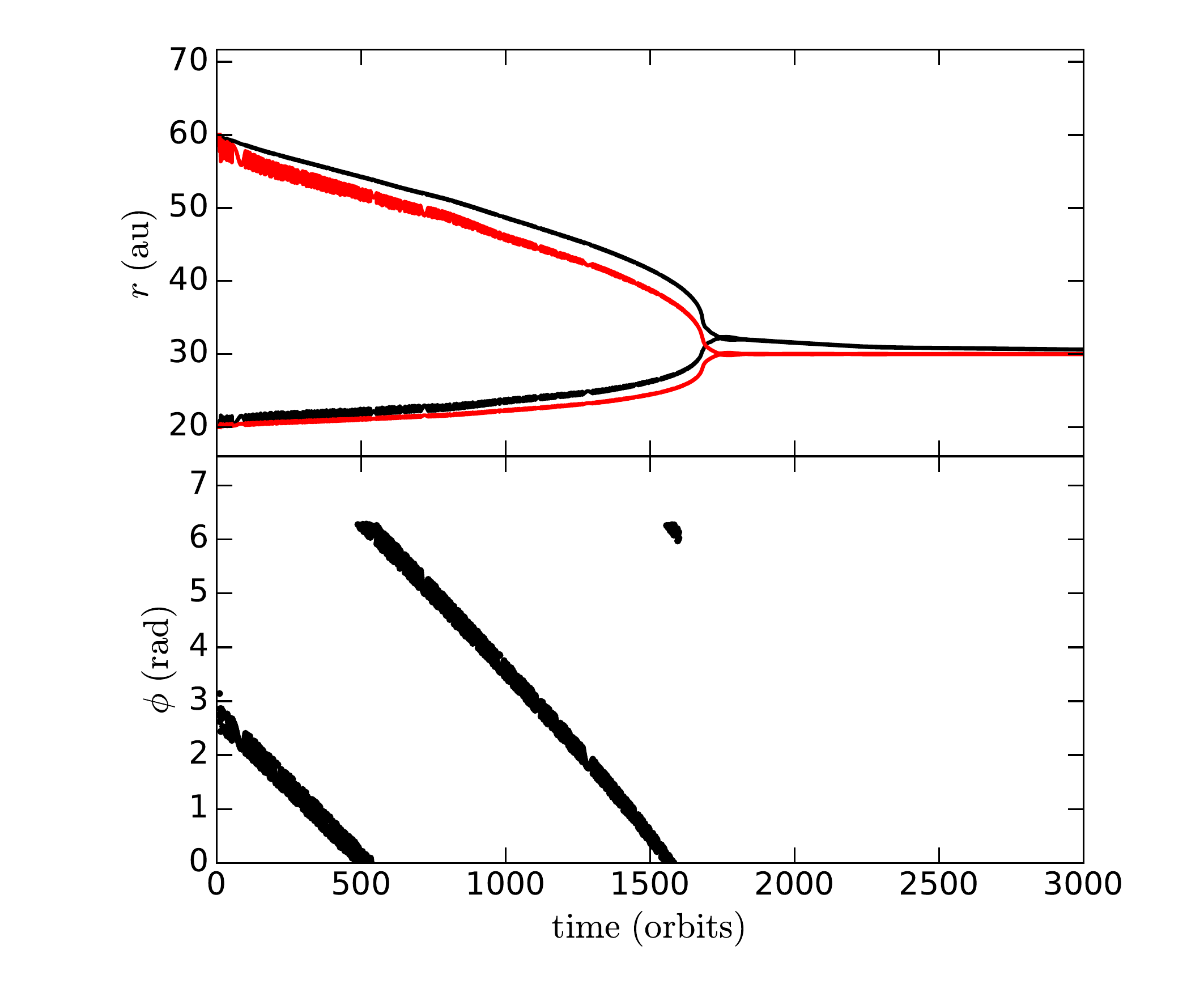}
\includegraphics[width=0.45\textwidth]{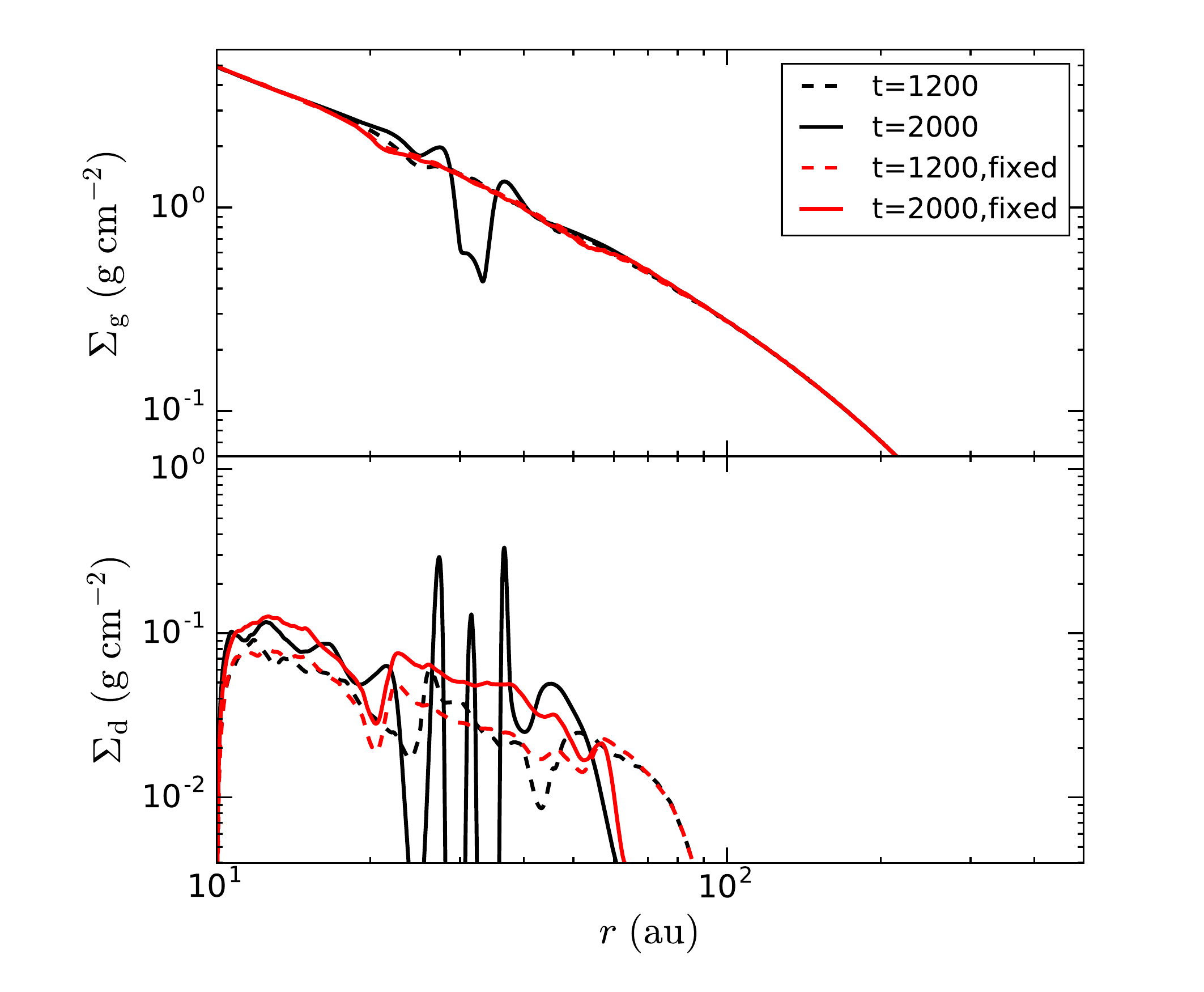}
\end{center}
\caption{Model with a low gas surface density (LowMg).  Upper left panel: planet's pericenter and apocenter as a function of time.  Red curves show the expected pericenter and apocenter if the angular momentum of the planet
was conserved. Lower left: the pericenter phase of the planet orbit as a function of time, which suggests a precession period of $\sim1000$ orbits. Right panels: azimuthally-averaged
gas (upper) and dust (lower) radial profiles at the stage of before circularization (1200 orbits) and after circularization (2000 orbits). For a comparison, we also show the corresponding radial profiles when the planet is fixed at the initial orbit as red lines.}\label{fig:dyn_LowMg}
\end{figure*}

\subsubsection{Disk mass}

We first discuss the effect of disk mass. As the planet circularization timescale is inversely proportional to the disk mass (e.g., \citealt{Papaloizou2000,Cresswell2007}, also confirmed by our simulations with a different disk mass but with a fixed eccentricity, which is not shown here),
we adopt a lower initial eccentricity to allow the faster planet circularization
when the disk mass is lower, labeled as model LowMg in Table~\ref{tab:para}.

We show the circularization process in the left panels of Figure~\ref{fig:dyn_LowMg}. The red lines in the upper left panel is the expected pericenter and apocenter radii if the angular momentum of the planet is conserved. We can see that the angular momentum gaining is not so prominent as in our fiducial model, which can be understood from Equation~(\ref{eq:Jpdot}) in the case of a low initial eccentricity. The circularization radius is thus very close to $\sim a_{\rm p}(1-e_{\rm p}^{2})=30\ {\rm au}$. The precession timescale shown in the lower left panel of Figure~\ref{fig:dyn_LowMg} becomes much longer compared with our fiducial model, which is related to the low disk mass\footnote{We find that the precession timescale is quite insensitive to the initial planetary eccentricity, e.g., $e_{\rm p}=0.5\sim0.8$.}. The planet-disk interaction induced planetary precession has been studied for different disk models \citep{Fontana2016,Davydenkova2018,Sefilian2019}.
More detailed comparison with these models will be carried out in future studies.

\begin{figure}[htbp]
%\vbox to3.2in{\rule{0pt}{3.2in}} \special{psfile=fig1.EPS voffset=0 hoffset=0 vscale=80 hscale=80 angle=0}
%\centering
\begin{center}
\includegraphics[width=0.45\textwidth]{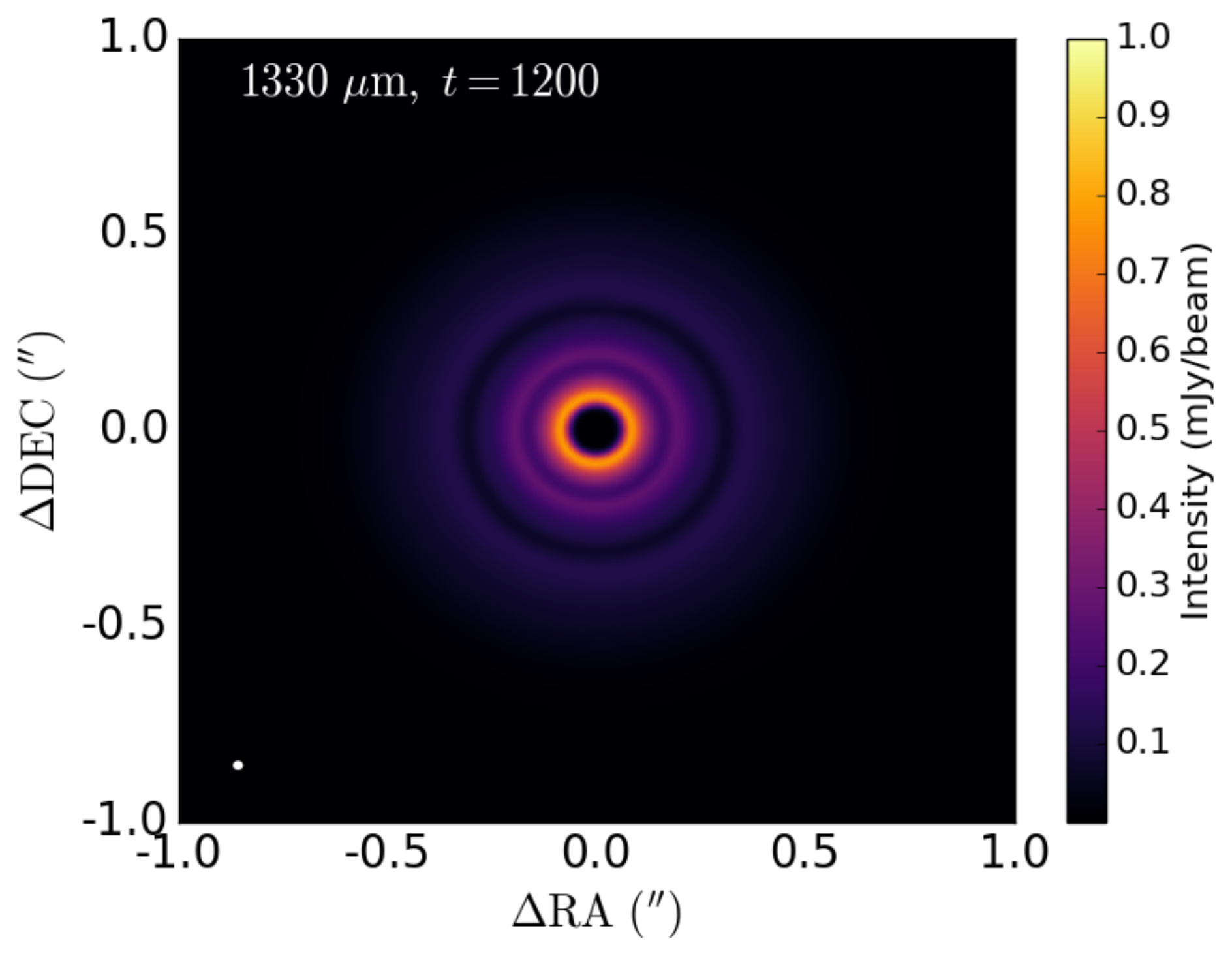}
\includegraphics[width=0.45\textwidth]{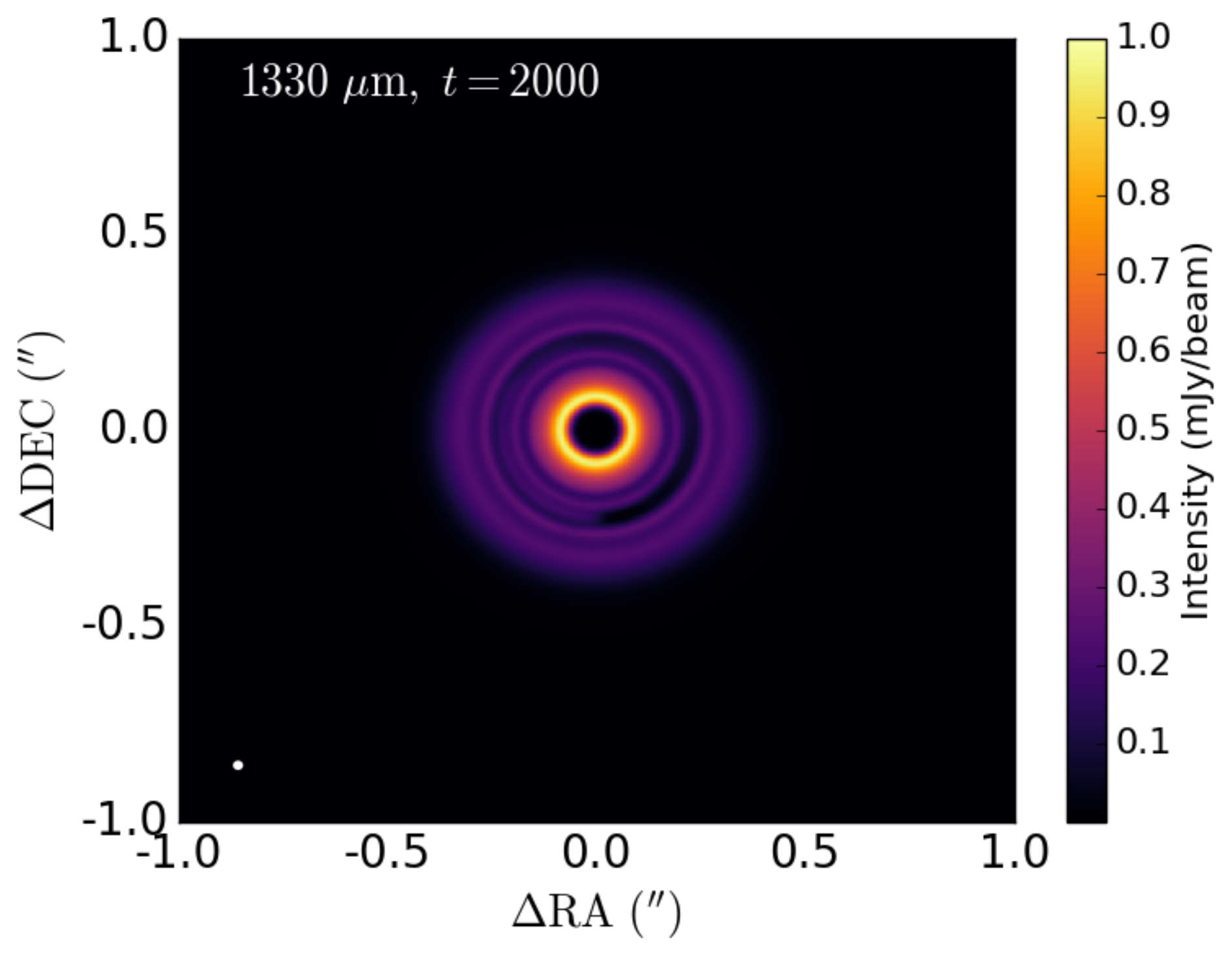}
\end{center}
\caption{Same as Figure\ \ref{fig:image} but for model LowMg. The planet is circularized at $\sim1600$ orbits, so $t=1200$ and $t=2000$ corresponds to the stages before and after the planet circularization. Note that dust
rings/gaps are formed both before and after the circularization.}\label{fig:image_LowMg}
\end{figure}

In the LowMg case, the dust disk becomes significantly smaller because of the
larger Stokes number of the dust as
defined in Equation~(\ref{eq:st}), which leads to a fast radial drift for the dust.
Due to the reduction of the radial drift timescale and the increase of the circularization
timescale with decreasing disk gas mass, we would expect that the dust radially drift inside
the planet circularization radius when we decrease the disk mass further.
In this case, a very compact disk
without any ringed-structures would be expected.

The gas sub-structure in the LowMg case can lead to a more prominent dust gap/ring
as shown in Figure~\ref{fig:dyn_LowMg} due to the rapid accumulation of the dust with the fast radial drift.
Three rings can be seen around $\sim27$, $\sim31$, and $\sim36$ au,
respectively, as shown in Figure~\ref{fig:dyn_LowMg}.
The rings at $27$ au and $36$ au are also close to the 3:2 resonance pair.

More interestingly, in the LowMg case, the dust rings/gaps can form {\it before} the
planet circularization as shown in the right panels of Figure~\ref{fig:dyn_LowMg}, different from other models we studied here.
We show the dust emission image at 1200 orbits in the upper panel of Figure~\ref{fig:image_LowMg}.
There is an outer gap located at 43 au and an inner one located at 25 au,
close to the current apocenter and pericenter of the orbit, respectively.
We find that such a gap opening process before orbit's circularization
are mainly related to the less massive disk and low disk viscosity. 
This is due to the density wave damping in the low viscosity disk. The dependence of the density wave launched by the planet on the Toorme Q parameter results in a much more efficient gap opening process when the Q parameter of the disk increases for a lower disk mass \citep{Rafikov2002,Li2009}. A lower disk viscosity is also essential in making a longer refilling timescale for the gas and producing deeper dust rings as we will show later. In addition, the tidal force on the disk becomes strongest at the pericenter and apocenter locations \citep{Muto2011}.  
All of these eventually produce the gaps (and rings) at the current apocenter and pericenter locations seen in the top panel of Figure~\ref{fig:image_LowMg}. The locations of those gaps/rings evolve with time as the planet circularizes.
This gap-opening process does not depend on the planet initial eccentricity
too sensitively so long as the initial $e_{\rm p}$ is high, as tested by our simulations with different initial eccentricities\footnote{The models with the same disk mass but with different initial eccentricities ($e_{\rm p}=0.5\sim0.8$) can produce gaps/rings structures at the pericenter/apocenter locations.}, so long as the viscous timescales 
are much longer than the gap opening time scale.  Since the planet can be circularized at a timescale of $\sim$Myr, which is comparable to the age of most protoplanetary disks, such evolution stage could be caught observationally. The gap opening before the planet circularization cautions against using the gap and ring location to infer the current planetary orbital radius.

To quantify the role of planet dynamics in the gap opening process, we test the case with the planet fixed on its initial eccentric orbit. The radial profiles of gas and dust are shown in the right panels of Figure~\ref{fig:dyn_LowMg}. Compared with the profiles at 1200 orbits in the LowMg model, the gap/rings are also present around the pericenter, although the features at the apocenter become weaker. This again indicates that the gap opening process is mainly controlled by the low disk mass (equivalently, a larger Toomre Q parameter) and the low viscosity (\citealt{Rafikov2002,Li2009}). An eccentric orbit produces an inefficient gap opening process due to the weak dynamical friction force in the high $e_{\rm p}$ regime \citep{Muto2011}.

As these processes all depend on how the angular momentum flux is transported throughout
the disk, the effect due to an adiabatic EoS versus locally isothermal condition is investigated.
\citet{Miranda2019} recently found that the locally isothermal EoS is inaccurate to capture rings/gaps
features under certain conditions, including when the the disk is nearly inviscid and the planet mass
is small (less than thermal mass). Here, we find out that the inaccurate treatment from the locally isothermal
EoS does not produce any significant differences on our main results.
The details are described in  Appendix~\ref{sec:app_ad}.

\begin{figure*}[htbp]
%\vbox to3.2in{\rule{0pt}{3.2in}} \special{psfile=fig1.EPS voffset=0 hoffset=0 vscale=80 hscale=80 angle=0}
%\centering
\begin{center}
\includegraphics[width=0.45\textwidth]{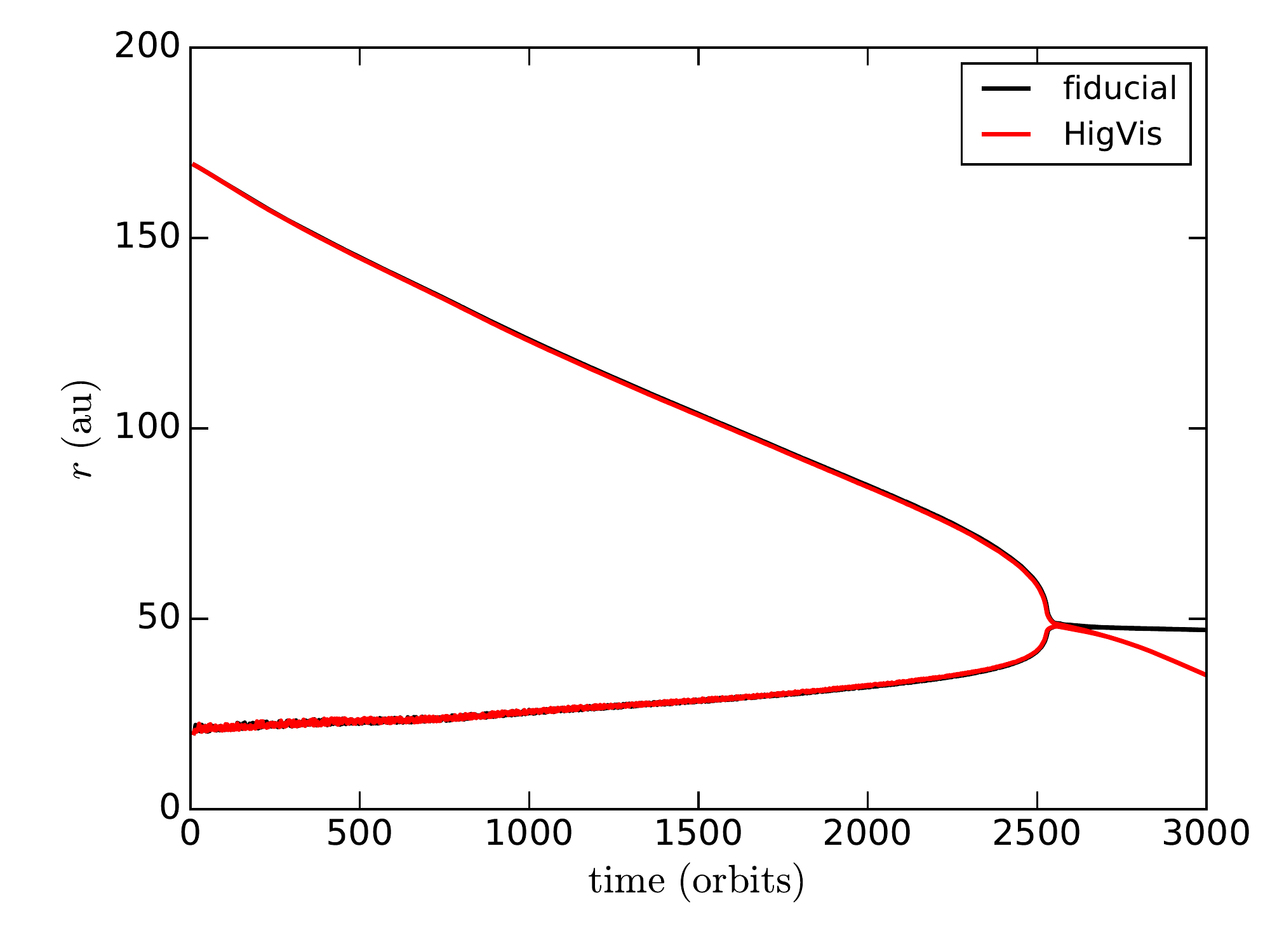}
\includegraphics[width=0.45\textwidth]{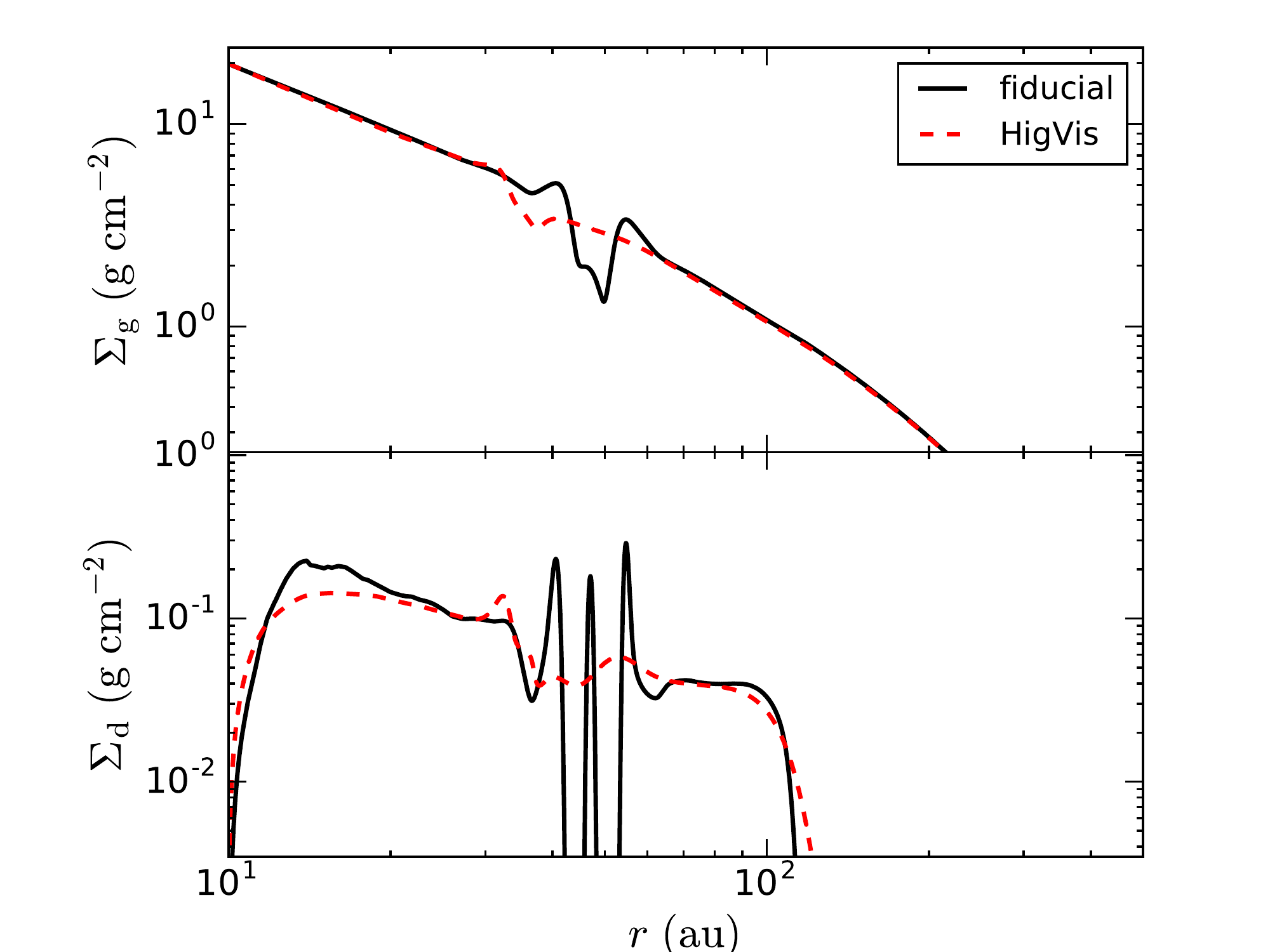}
\end{center}
\caption{Models with different disk viscosities.  Left: planet's pericenter and
apocenter as a function of time. The model with a high viscosity (HigVis) shows a
similar circularization process with our fiducial model. Right: Azimuthally-averaged
gas (upper) and dust (lower) radial profiles. All radial distributions are shown at 3000 orbits when the planet is circularized.}\label{fig:dyn_HigVis}
\end{figure*}

\subsubsection{Disk viscosity}

We then consider the model HigVis with a higher $\alpha_{\rm vis}$.
The planet dynamics (circularization radius, timescale) is almost the same as
our fiducial model as shown in Figure~\ref{fig:dyn_HigVis}. After the planet's circularization,
the migration rate significantly increases, different from the very slow migration with the
smaller viscosity as shown in Figure~\ref{fig:circ}. This is also consistent with previous
works by \citet{Li2009} and \citet{Yu2010}. The fast inward migration of the planet
causes the gaps/rings move closer to the central region after a longer evolution time.
A smaller gap depth is also attributed to the high viscosity \citep{Fung2014,Kanagawa2015,Zhu2019}.
A shallower gap of the gas distribution for the high viscosity model leads to the
dust gap and ring being shallow as well, which further makes the detection of such gap/rings harder.

\subsubsection{Dust sizes}\label{sec:dustsize}

\begin{figure}[htbp]
%\vbox to3.2in{\rule{0pt}{3.2in}} \special{psfile=fig1.EPS voffset=0 hoffset=0 vscale=80 hscale=80 angle=0}
%\centering
\begin{center}
\includegraphics[width=0.45\textwidth]{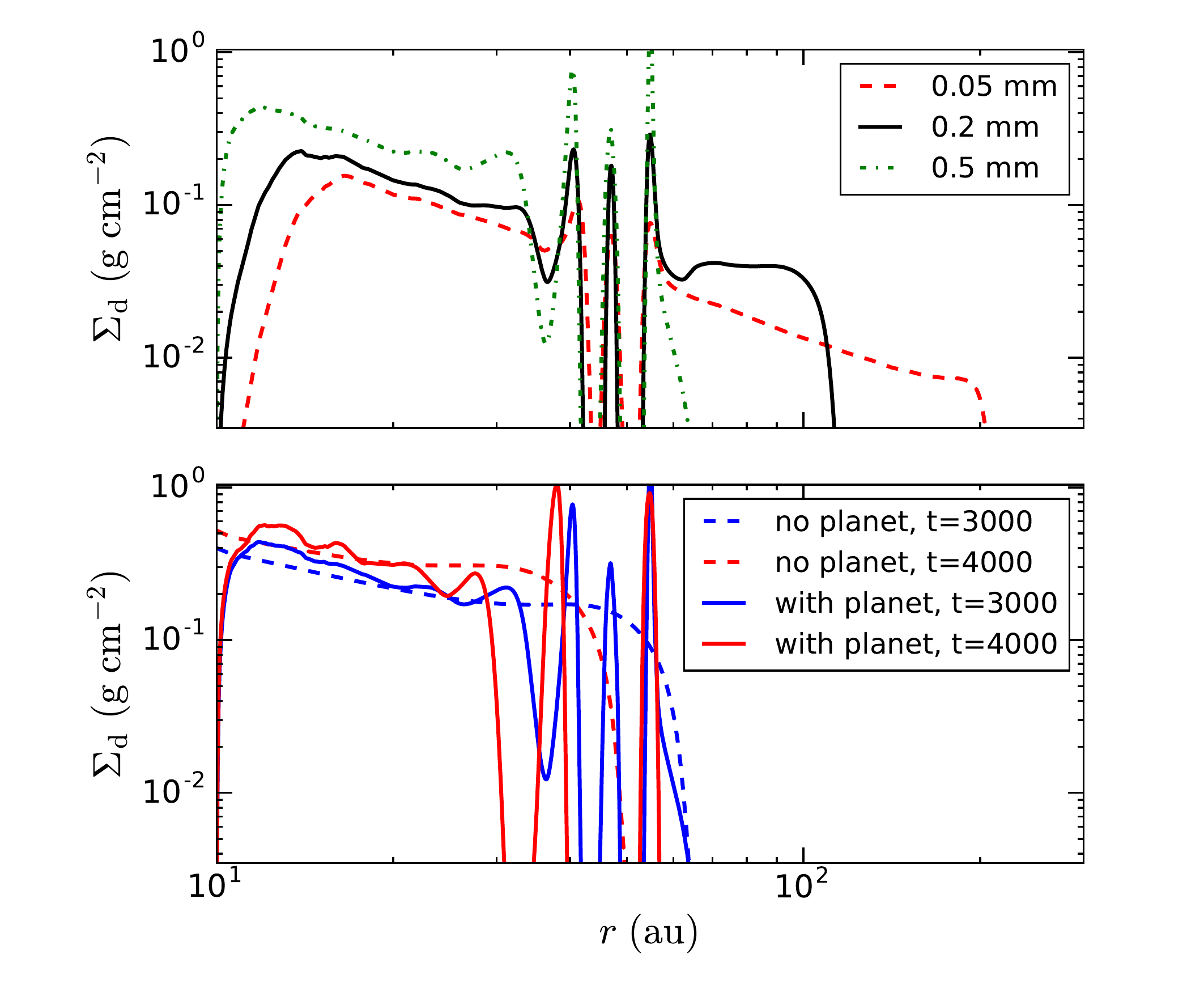}
\end{center}
\caption{The effect of dust sizes on the ring formation. Upper: Azimuthally-averaged
dust radial profiles at $t = 3000$ orbits
with different dust sizes based on Model S2 (red), fiducial (black) and
S3 (green), respectively. Lower: Effect of a planet
on the radial drift of the large-sized dust. The planet case corresponds to
model S3.}\label{fig:dyn_size}
\end{figure}

We find that the incorporation of dust has almost no impact on planet dynamics,  while the dust size is a factor that can influence the dust radial drift. We can, therefore, explore the dust ring formation by adopting different dust sizes.
We compare the dust distributions for $s_{\rm d}=0.05\ {\rm mm}$ and
$s_{\rm d}=0.5\ {\rm mm}$ with our fiducial model in Figure~\ref{fig:dyn_size}, where dust can only be significantly perturbed after the planet circularization.

We find that the overall size of the dust disk gradually gets smaller when the dust size increases from
$s_{\rm d}=0.05\ {\rm mm}$ (red) to $s_{\rm d}=0.5\ {\rm mm}$ (green),
although the gas distribution remains the same.
For all models, the dust gap/rings appear in the same location due to the same planet circularized radius,
though the dust concentration in rings as compared to the surrounding regions can vary.
They all have three rings similar with our fiducial model with $s_{\rm d}=0.2\ {\rm mm}$.

In particular, for Model S3 with $s_{\rm d}=0.5\ {\rm mm}$, the dust drifts radially
quicker than our fiducial model but the dust radial drift is almost halted at the planet's
circularization radius. As
shown in the lower panel of Figure~\ref{fig:dyn_size}, the outer dust ring at 1.4 Myr
is almost the same as that at 1.1 Myr. In contrast, such dust would have drifted significantly
inward between $1.1$ and $1.4$ Myr if there is no planet (shown as the dash lines in
the bottom panel of Figure~\ref{fig:dyn_size}).
This suggests that, depending on the interplay among several timescales for
the dust drift, planet orbital circularization, and gap opening processes,
the circularized planet orbit can effectively stop the dust drift at large disk radii,
accompanied with relatively narrow, dense dust rings.

\section{Conclusions}\label{sec:conclusions}

In this work, we have performed 2D high-resolution hydrodynamical simulations with
LA-COMPASS \citep{Li2005,Li2009,Fu2014} to study the interaction of a highly
eccentric super-Earth with a dusty gaseous disk.  The highly eccentric super-Earth initialized
in our simulations is the main assumption in this work, which is motivated by the scenario that many planets can form at small radii but some can be scattered into highly eccentric orbits. One single dust species is implemented
in gas-dust disk dynamics with self-consistent feedback. Disk self-gravity is also included.
With the long time evolution of the coupled gas/dust and the embedded planet,
we study the planet circularization process, its consequent impact on the dust dynamics,
and the observational features as well. We also explore how different disk mass,
viscosity, planetary parameters and dust size can influence the appearance of dust rings.

We confirm the parameter dependence of circularization timescale on the initial eccentricity, disk mass, and disk scale height in previous works \citep{Papaloizou2000,Cresswell2008,Muto2011,Ida2019}.  A higher viscosity can speed up the migration rate significantly after the planet's circularization, but has no influence on the planetary dynamics before its circularization. The inclusion of dust has almost no impact on the planetary circularization and migration processes. Our main findings are summarized as follows, mostly applicable to situations
with initially high orbital eccentricity:

\begin{itemize}

\item The planet's orbit gains angular momentum when the initial planetary eccentricity is high enough, consistent with the results using analytic analysis and N-body simulations. This results in a large circularization orbital radius for the planet. The planet opens up a partial gap in gas, producing dust gaps and rings
around the circularization radius after planet's circularization. This could explain the gaps/rings observed at the outer region of disks. Before the circularization or when we fix the planet at its initial eccentric orbit, the disturbance of the planet on the dusty disk is very week when the disk mass is high ($\sim0.02\ M_{\odot}$).

\item When the disk mass is low ($\sim 5\times 10^{-3} M_{\sun}$), however,
we find that gaps and rings can be produced at the current pericenter and apocenter location even before its orbital circularization for a low disk viscosity. This is also true when we fix the planet at the initial eccentric orbit.
This cautions against using the gap and ring location to infer the current planetary
orbital radius.

\item An eccentric planet can potentially slow down the dust radial drift in the outer region of the disk,
particularly when the planet's orbit is circularized faster than the radial drift timescale and
a low disk viscosity that slows down the planetary inward migration. The dust drift is stopped at the
rings with relatively little emission from the surrounding regions. Observations at longer wavelengths tracing the large dust particles will be particularly useful in constraining this configuration.

\end{itemize}

We have explored a relatively straightforward limit with a single super-Earth planet on an eccentric orbit.
Configurations with multiple planets (perhaps with a mixture of massive and smaller planets) will also
be quite interesting. As some of the observed protoplanetary disks are relatively old with possibly low
total disk mass, the circularization timescales for such planets can be long, exceeding $1$ Myr.
It is then imperative to include these processes when interpreting the origin and mechanisms for
dust ring and gap formation.

\acknowledgments
We thank the referee for very helpful comments to improve the presentation of the paper. YPL, HL and SL gratefully acknowledge the support by LANL/CSES and NASA/ATP.
Simulations of this work were performed with LANL Institutional Computing resources.
HL gratefully acknowledges useful discussions with S. Ida, R. Miranda and R. Rafikov.

\software{\texttt{Astropy} \citep{Astropy2013},
          \texttt{LA-COMPASS} \citep{Li2005,Li2009},
          \texttt{Matplotlib} \citep{Hunter2007},
          \texttt{Numpy} \citep{vanderWalt2011},
          \texttt{RADMC-3D} \citep{Dullemond2012}
          }

\appendix

\section{The effect of adiabatic disks}\label{sec:app_ad}

Recently, \citet{Miranda2019} found that the locally isothermal simulations tend to overestimate
the contrast of rings/gaps, and even mis-identify the planet position that is responsible those features.
To quantify this effect on our simulation results, we choose the run LowMg to compare with another run where we incorporate the energy equation with an adiabatic index $\gamma=1.001$ as in \citet{Miranda2019},
while keeping all other parameters being the same.
We do not implement dust in the adiabatic run as in \citet{Miranda2019},
since we found that dust has no impact in the planet dynamics and the feedback effect
on the gas is small.

\begin{figure}[htbp]
%\vbox to3.2in{\rule{0pt}{3.2in}} \special{psfile=fig1.EPS voffset=0 hoffset=0 vscale=80 hscale=80 angle=0}
%\centering
\begin{center}
\includegraphics[width=0.33\textwidth]{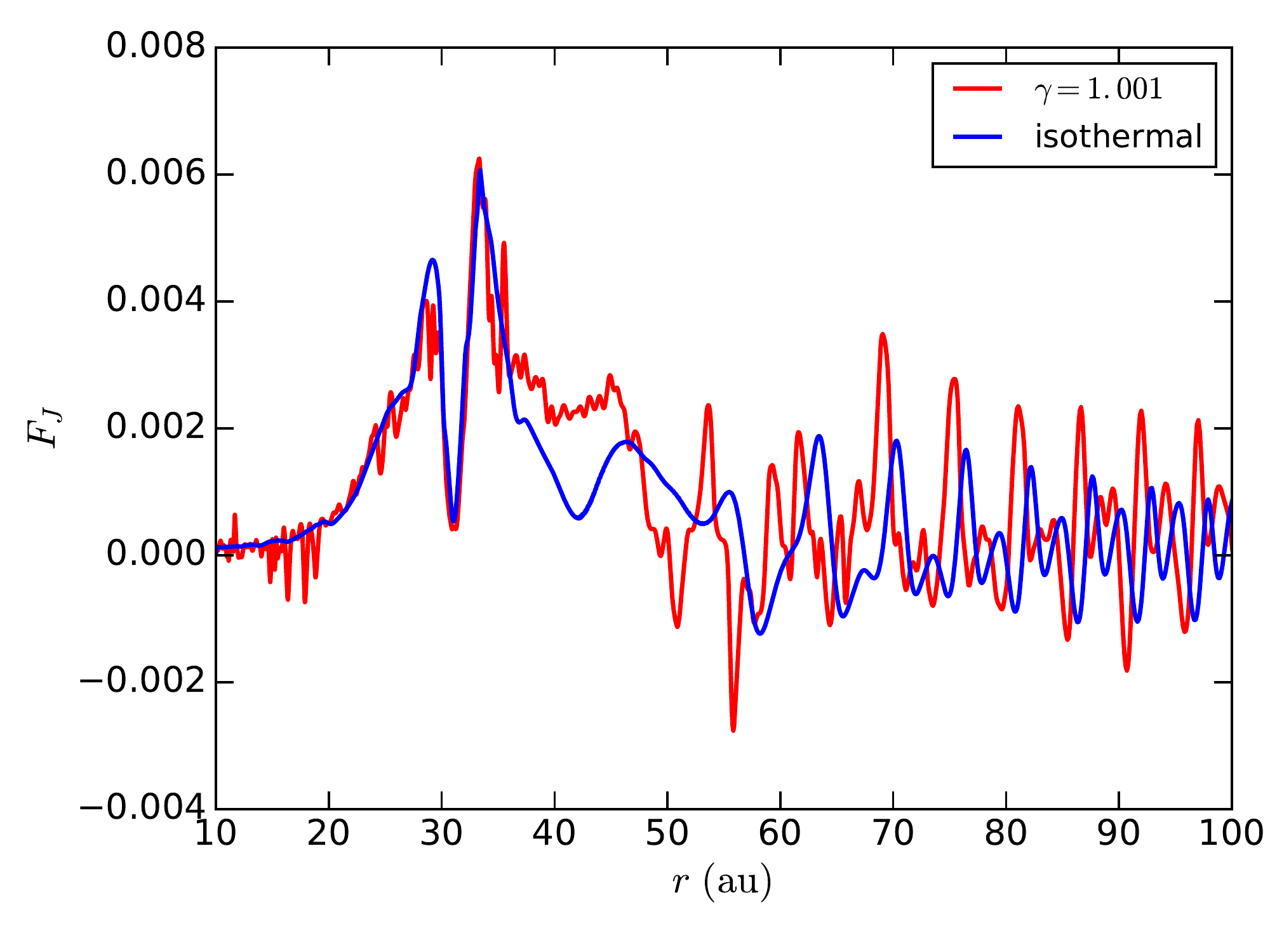}
\includegraphics[width=0.33\textwidth]{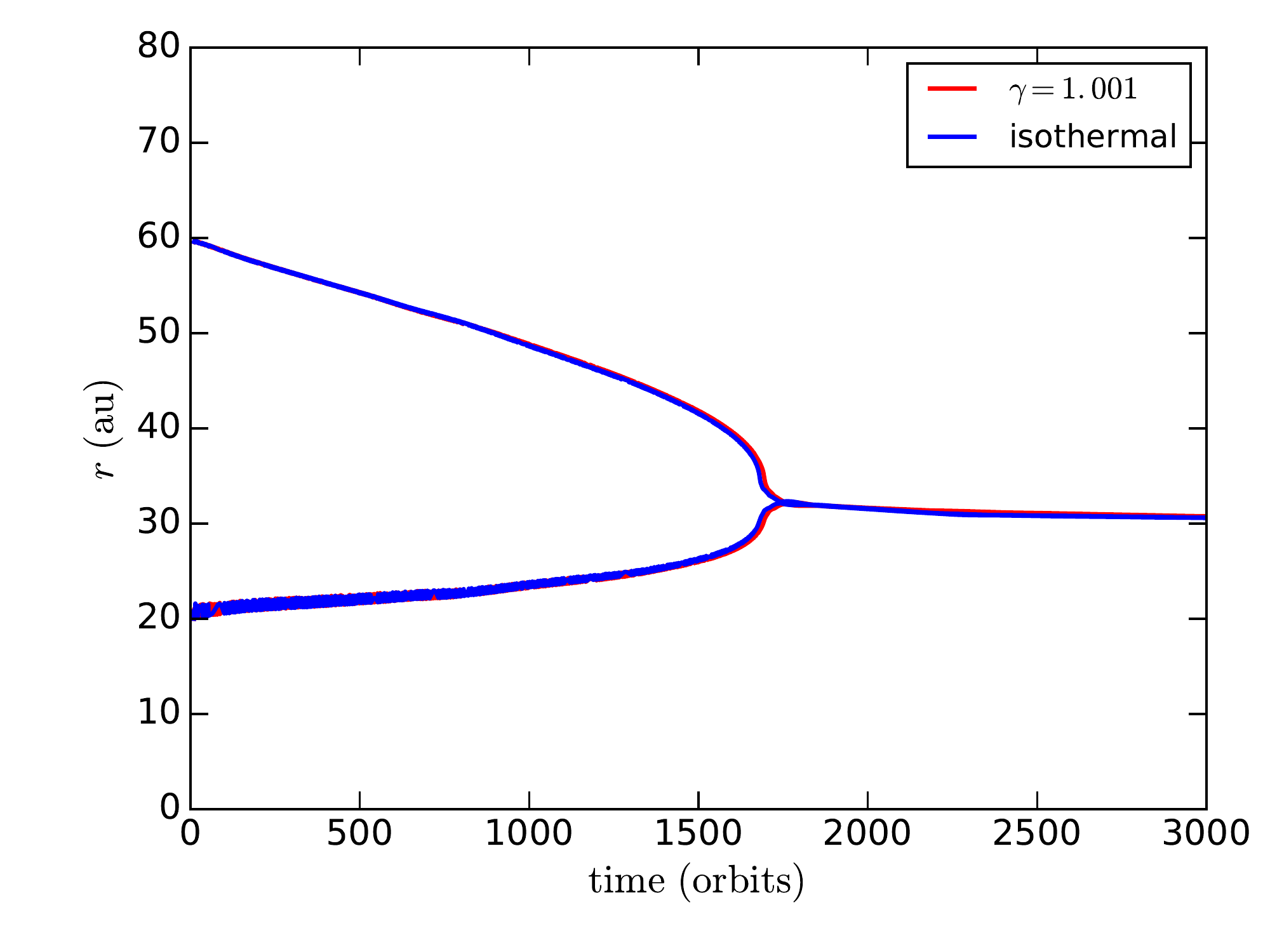}
\includegraphics[width=0.33\textwidth]{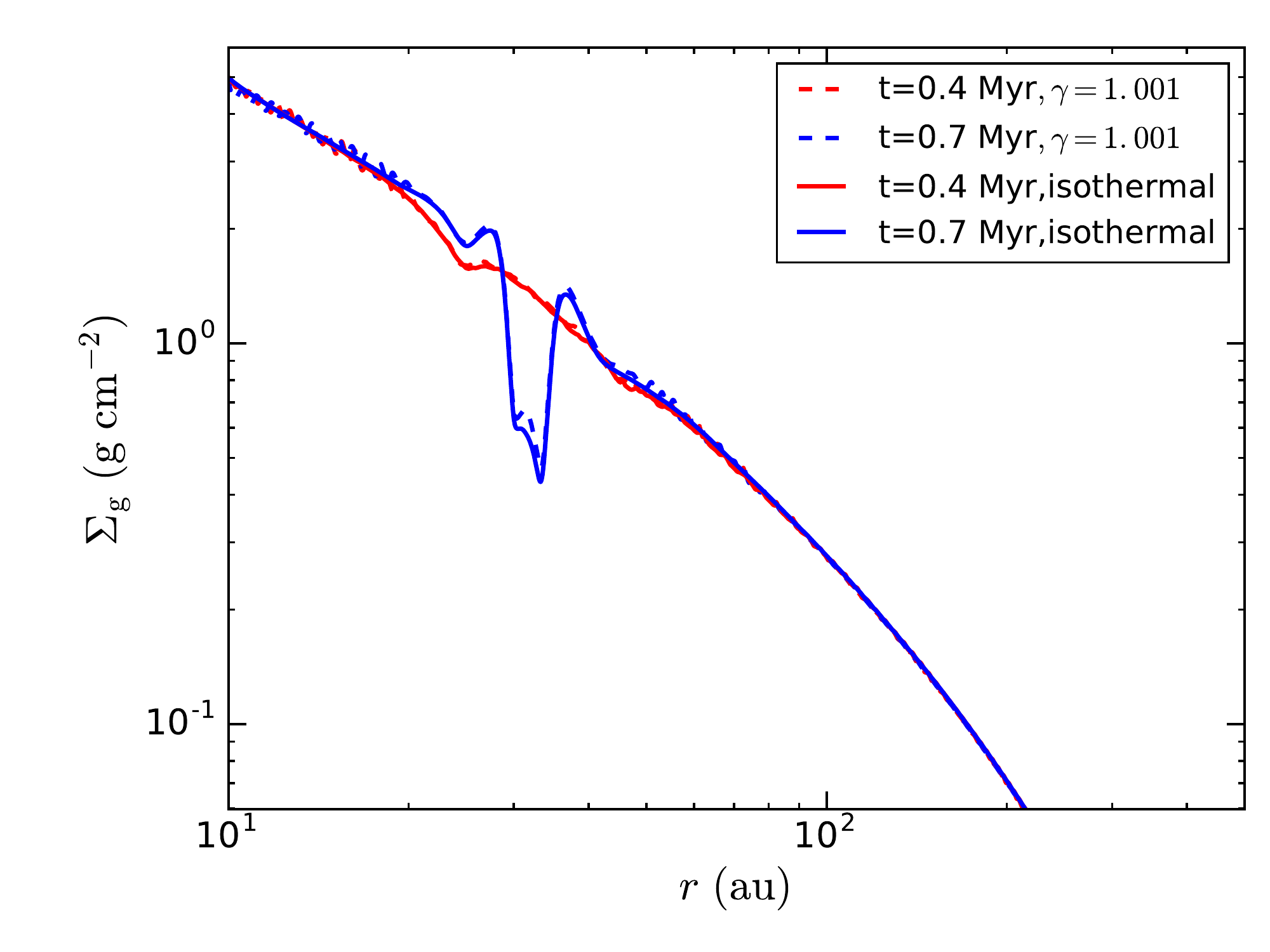}
\end{center}
\caption{The effect of adiabatic EoS ($\gamma=1.001$) vs. the locally isothermal LowMg run on the planet dynamics and
gas surface density profiles. Left: radial profiles of the planet-induced angular momentum flux in the code unit at 2000 orbits.
Middle: the planetary orbital apocenter and pericenter as a function of time. The isothermal and
adiabatic with $\gamma=1.001$ runs are almost indistinguishable. Right: the radial distributions of the
gas surface density at two times ($t = $1200 and 2000 orbits) for two runs.
Again, the difference is small.}\label{fig:LowMg_ad}
\end{figure}

In Figure~\ref{fig:LowMg_ad}, we show the radial profile of the integrated angular momentum flux (AMF)
at 2000 orbits (left), and the planet circularization process (middle), and the radial distribution of
gas surface density (right) at two different times (corresponding to 1200 and 2000 orbits)
for our locally isothermal case and the adiabatic one. The discrepancy of AMF is only marginal
for the region where the planet is circularized ($\sim 30\ {\rm au}$). The offset beyond $r=60\ {\rm au}$
could be the imprint of the planetary circularization process, which, however, does not leave
significant features on the gas. This is actually consistent with the expectation from \citet{Miranda2019}.
When the planet mass is $M_{\rm p}=1\ {M_{\rm th}}$, where the thermal mass
$M_{\rm th}=h_{\rm 0,p}^{3}M_{\star}\simeq7\ {M_{\oplus}}$ (here we measure the disk scale
height at the initial semi-major axis $a_{\rm p}$), the error caused by the locally isothermal case is not significant.
Furthermore, we find that the circularization timescale and the circularization radius are almost the
same for these two case, as shown in the middle panel of Figure~\ref{fig:LowMg_ad}.
Such an indistinguishable planet dynamics leads to the similar gas surface density profile
as shown in the right panel of  Figure~\ref{fig:LowMg_ad}. Therefore, we should also expect
that the impact on the dust distribution in terms of the dust ring appearance and their
locations should be insignificant if we use a similar adiabatic EoS.

\section{Parameter dependence of planet circularization timescale}\label{sec:app_tcirc}

Here we further explore the non-linear dependence of the planet circularization timescale on the
disk scale height (or sound speed).

\begin{figure}[htbp]
%\vbox to3.2in{\rule{0pt}{3.2in}} \special{psfile=fig1.EPS voffset=0 hoffset=0 vscale=80 hscale=80 angle=0}
%\centering
\begin{center}
\includegraphics[width=0.45\textwidth]{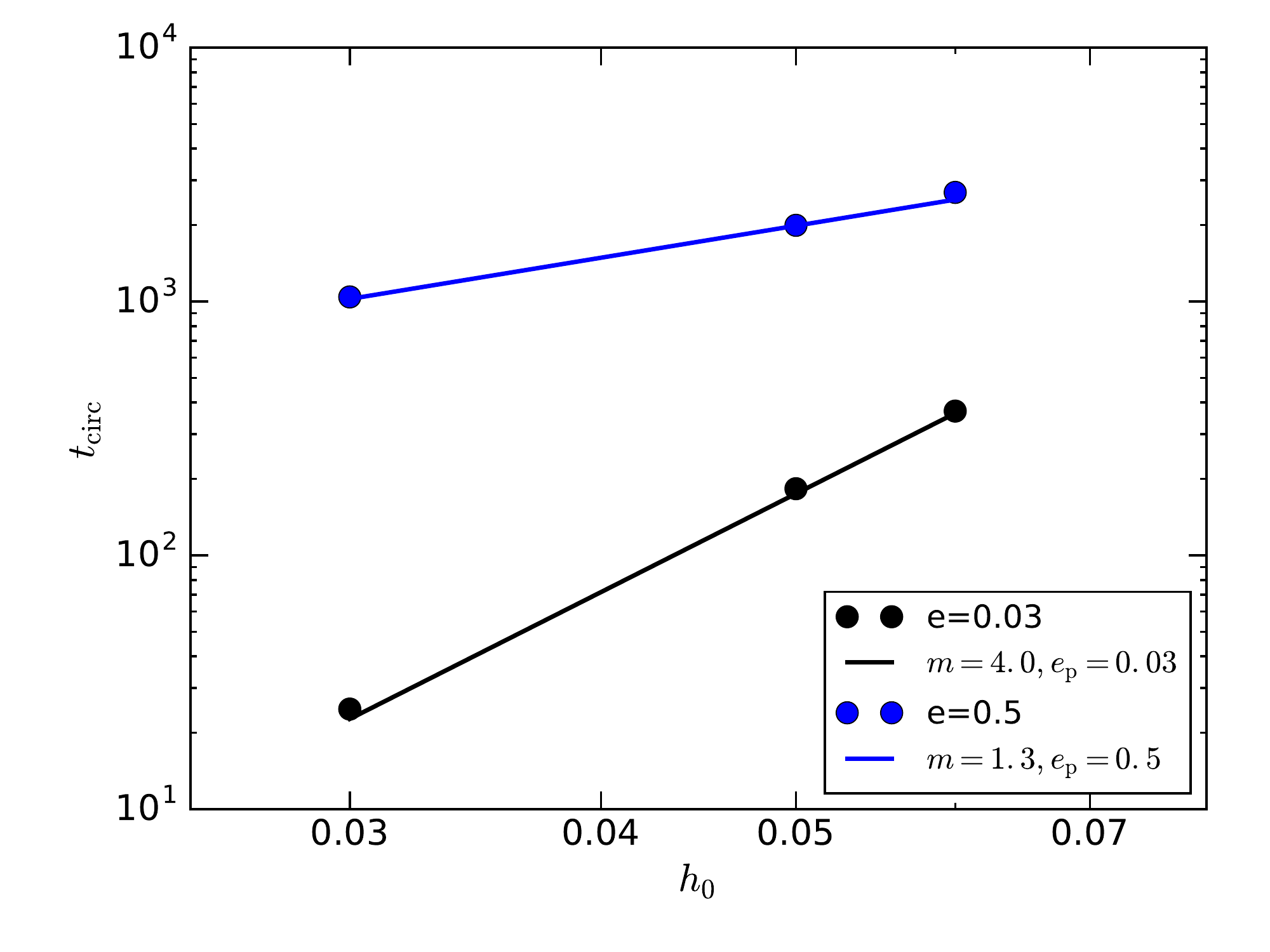}
\end{center}
\caption{Parameter dependence of circularization timescale ($t_{\rm circ}$) on the disk scale height $h_{0}$ at $r_{0}$ (or the sound speed). We use a simple power-law model to quantify the dependence of $t_{\rm circ}$ on $h_{0}$ in the high $e_{\rm p}$ and low $e_{\rm p}$ regimes.}\label{fig:circ_all}
\end{figure}

Since the inclusion of dust does not change the dynamics of the planet,
we only consider a pure gaseous disk without dust. Our default disk gas mass
is $2\times10^{-2}\ M_{\odot}$ with the same radial distribution as in Equation~(\ref{eq:gas}).
The other default parameters are $M_{\rm p}=10\ M_{\oplus}$, and a softening factor of $0.7h_{\rm g}$
as our fiducial model. The planetary semi-major axis is $r_{\rm p}=2.0$ in code unit with $r_{0}=50\ {\rm au}$.
We explore two regimes, one is $e_{\rm p}\gg c_{\rm s,0}$ (i.e., the supersonic case),
where $c_{\rm s,0}$ is the sound speed at the initial planetary semi-major axis
($c_{\rm s,0}=r_{\rm p}^{0.25}h_{0}$), and the other one is $e_{\rm p}\lesssim c_{\rm s,0}$
(i.e., the subsonic case).
For the supersonic case, we fit the time evolution profile of the eccentricity as
shown in the middle panel of Figure~\ref{fig:circ} with $de_{\rm p}/dt= -Ae_{\rm p}^{-m}$.
The circularization timescale is then $t_{\rm circ}=e_{\rm p}^{\alpha+1}/A(\alpha+1)$. For the subsonic case,
it has been suggested that $t_{\rm circ}$ is independent of the initial eccentricity $e_{\rm p}$
\citep{Papaloizou2000}, so we fit the temporal eccentricity
profile for the very low eccentricity part ($e_{\rm p}\ll h_{0}$) with $de_{\rm p}/dt= -Ae_{\rm p}$. And we define the corresponding circularization timescale $t_{\rm circ}=1/A$. We have six runs with three different $h_{0}$ as shown in Figure~\ref{fig:circ_all} for two regimes.

We use a single power-law model $t_{\rm circ}(h_{0})\propto\ h_{0}^{-m}$
to quantify the dependence of $t_{\rm circ}$ on $h_{0}$.
We find that the power-law index is $\alpha\sim1.3$ for the low eccentricity regime (subsonic case),
while $\alpha\sim4.0$  for the high eccentricity regime (supersonic case), close to the expectation of the functional form from \citet{Papaloizou2000} and \citet{Ida2019}.
The slight difference of $m$ in the high eccentricity regime could be due to the softening effect and the transition average between supersonic and subsonic regimes.

\end{CJK*}


\begin{thebibliography}{}

\bibitem[ALMA Partnership et al.(2015)]{ALMA2015} ALMA Partnership, Brogan, C.~L., P{\'e}rez, L.~M., et al.\ 2015, \apjl, 808, L3

%\bibitem[Andrews(2015)]{Andrews2015} Andrews, S.~M.\ 2015, \pasp, 127, 961

\bibitem[Andrews et al.(2018)]{Andrews2018} Andrews, S.~M., Huang, J., P{\'e}rez, L.~M., et al.\ 2018, \apj, 869, L41.

%\bibitem[Andrews et al.(2010)]{Andrews2010} Andrews, S.~M., Wilner, D.~J., Hughes, A.~M., Qi, C., \& Dullemond, C.~P.\ 2010, \apj, 723, 1241

\bibitem[Andrews et al.(2016)]{Andrews2016} Andrews, S.~M., Wilner, D.~J., Zhu, Z., et al.\ 2016, \apjl, 820, L40

\bibitem[Ansdell et al.(2018)]{Ansdell2018} Ansdell, M., Williams, J.~P., Trapman, L., et al.\ 2018, \apj, 859, 21

%\bibitem[Armitage(2010)]{Armitage2010} Armitage, P.~J.\ 2010, Astrophysics of Planet Formation, by Philip J.~Armitage, 294 pp.~ISBN %978-0-521-88745-8 (hardback).~Cambridge, UK: Cambridge University Press, 2010.,

%\bibitem[Armitage(2011)]{Armitage2011} Armitage, P.~J.\ 2011, \araa, 49, 195

\bibitem[Arzamasskiy et al.(2018)]{Arzamasskiy2018} Arzamasskiy, L., Zhu, Z., \& Stone, J.~M.\ 2018, \mnras, 475, 3201

\bibitem[Astropy Collaboration et al.(2013)]{Astropy2013} Astropy Collaboration, Robitaille, T.~P., Tollerud, E.~J., et al.\ 2013, \aap, 558, A33

\bibitem[Birnstiel et al.(2010a)]{Birnstiel2010a} Birnstiel, T., Dullemond, C.~P., \& Brauer, F.\ 2010a, \aap, 513, A79

%\bibitem[Birnstiel et al.(2018)]{Birnstiel2018} Birnstiel, T., Dullemond, C.~P., Zhu, Z., et al.\ 2018, \apjl, 869, L45

\bibitem[Bitsch et al.(2013)]{Bitsch2013} Bitsch, B., Crida, A., Libert, A.-S., \& Lega, E.\ 2013, \aap, 555, A124

\bibitem[Boehler et al.(2018)]{Boehler2018} Boehler, Y., Ricci, L., Weaver, E., et al.\ 2018, \apj, 853, 162

\bibitem[Cieza et al.(2017)]{Cieza2017} Cieza, L.~A., Casassus, S., P{\'e}rez, S., et al.\ 2017, \apjl, 851, L23

\bibitem[Clarke et al.(2018)]{Clarke2018} Clarke, C.~J., Tazzari, M., Juhasz, A., et al.\ 2018, \apj, 866, L6.

\bibitem[Cox et al.(2017)]{Cox2017} Cox, E.~G., Harris, R.~J., Looney, L.~W., et al.\ 2017, \apj, 851, 83

\bibitem[Cresswell et al.(2007)]{Cresswell2007} Cresswell, P., Dirksen, G., Kley, W., \& Nelson, R.~P.\ 2007, \aap, 473, 329

\bibitem[Cresswell, \& Nelson(2008)]{Cresswell2008} Cresswell, P., \& Nelson, R.~P.\ 2008, \aap, 482, 677

\bibitem[D'Angelo et al.(2006)]{DAngelo2006} D'Angelo, G., Lubow, S.~H., \& Bate, M.~R.\ 2006, \apj, 652, 1698.

%\bibitem[Dawson, \& Chiang(2014)]{Dawson2014} Dawson, R.~I., \& Chiang, E.\ 2014, Science, 346, 212.

\bibitem[Davydenkova, \& Rafikov(2018)]{Davydenkova2018} Davydenkova, I., \& Rafikov, R.~R.\ 2018, \apj, 864, 74

\bibitem[Dipierro et al.(2018)]{Dipierro2018} Dipierro, G., Ricci, L., P{\'e}rez, L., et al.\ 2018, \mnras, 475, 5296

\bibitem[Dipierro et al.(2015)]{Dipierro2015} Dipierro, G., Price, D., Laibe, G., et al.\ 2015, \mnras, 453, L73

\bibitem[Dong et al.(2017)]{Dong2017} Dong, R., Li, S., Chiang, E., et al.\ 2017, \apj, 843, 127

\bibitem[Dong et al.(2018b)]{Dong2018b} Dong, R., Li, S., Chiang, E., et al.\ 2018b, \apj, 866, 110

\bibitem[Dong et al.(2018a)]{Dong2018a} Dong, R., Liu, S.-. yuan ., Eisner, J., et al.\ 2018a, \apj, 860, 124

\bibitem[Dong et al.(2015)]{Dong2015} Dong, R., Zhu, Z., \& Whitney, B.\ 2015, \apj, 809, 93.

\bibitem[Duffell \& Chiang(2015)]{Duffell2015} Duffell, P.~C., \& Chiang, E.\ 2015, \apj, 812, 94

\bibitem[Dullemond et al.(2012)]{Dullemond2012} Dullemond, C.~P., Juhasz, A., Pohl, A., et al.\ 2012, Astrophysics Source Code Library, ascl:1202.015

\bibitem[Fedele et al.(2018)]{Fedele2018} Fedele, D., Tazzari, M., Booth, R., et al.\ 2018, \aap, 610, A24

\bibitem[Fontana \& Marzari(2016)]{Fontana2016} Fontana, A., \& Marzari, F.\ 2016, \aap, 589, A133

\bibitem[Fu et al.(2014)]{Fu2014} Fu, W., Li, H., Lubow, S., Li, S., \& Liang, E.\ 2014, \apjl, 795, L39

\bibitem[Fung et al.(2014)]{Fung2014} Fung, J., Shi, J.-M., \& Chiang, E.\ 2014, \apj, 782, 88

\bibitem[Gonzalez et al.(2015)]{Gonzalez2015} Gonzalez, J.-F., Laibe, G., Maddison, S.~T., et al.\ 2015, \mnras, 454, L36.

\bibitem[Goldreich \& Sari(2003)]{Goldreich2003} Goldreich, P., \& Sari, R.\ 2003, \apj, 585, 1024

 \bibitem[Goldreich \& Tremaine(1978)]{Goldreich1978} Goldreich, P., \& Tremaine, S.~D.\ 1978, \icarus, 34, 240

\bibitem[Goldreich \& Tremaine(1980)]{Goldreich1980} Goldreich, P., \& Tremaine, S.\ 1980, \apj, 241, 425


%\bibitem[Huang et al.(2018a)]{Huang2018} Huang, J., Andrews, S.~M., Cleeves, L.~I., et al.\ 2018a, \apj, 852, 122

%\bibitem[Huang et al.(2018)]{Huang2018} Huang, J., Andrews, S.~M., Dullemond, C.~P., et al.\ 2018, \apj, 869, L42.

\bibitem[Hunter(2007)]{Hunter2007} Hunter, J.~D.\ 2007, Computing in Science and Engineering, 9, 90

\bibitem[Ida et al.(2019)]{Ida2019} Ida, S., et al. 2019, in preparation

\bibitem[Isella et al.(2016)]{Isella2016} Isella, A., Guidi, G., Testi, L., et al.\ 2016, Physical Review Letters, 117, 251101

\bibitem[Isella et al.(2018)]{Isella2018} Isella, A., Huang, J., Andrews, S.~M., et al.\ 2018, \apjl, 869, L49

\bibitem[Isella et al.(2009)]{Isella2009} Isella, A., Carpenter, J.~M., \& Sargent, A.~I.\ 2009, \apj, 701, 260

%\bibitem[Jin et al.(2016)]{Jin2016} Jin, S., Li, S., Isella, A., Li, H., \& Ji, J.\ 2016, \apj, 818, 76

\bibitem[Jin et al.(2016)]{Jin2016} Jin, S., Li, S., Isella, A., et al.\ 2016, \apj, 818, 76

\bibitem[Juri{\'c} \& Tremaine(2008)]{Juric2008} Juri{\'c}, M., \& Tremaine, S.\ 2008, \apj, 686, 603.

\bibitem[Kanagawa et al.(2015)]{Kanagawa2015} Kanagawa, K.~D., Muto, T., Tanaka, H., et al.\ 2015, \apjl, 806, L15

\bibitem[Kozai(1962)]{Kozai1962} Kozai, Y.\ 1962, \aj, 67, 591.

\bibitem[Lega et al.(2013)]{Lega2013} Lega, E., Morbidelli, A., \& Nesvorn{\'y}, D.\ 2013, \mnras, 431, 3494.

\bibitem[Li et al.(2001)]{Li2001} Li, H., Colgate, S.~A., Wendroff, B., \& Liska, R.\ 2001, \apj, 551, 874

%\bibitem[Li et al.(2000)]{Li2000} Li, H., Finn, J.~M., Lovelace, R.~V.~E., \& Colgate, S.~A.\ 2000, \apj, 533, 1023

\bibitem[Li et al.(2009)]{Li2009} Li, H., Lubow, S.~H., Li, S., \& Lin, D.~N.~C.\ 2009, \apjl, 690, L52

\bibitem[Li et al.(2005)]{Li2005} Li, H., Li, S., Koller, J., et al.\ 2005, \apj, 624, 1003

\bibitem[Li et al.(2009)]{Li2009} Li, S., Buoni, M.~J., \& Li, H.\ 2009, \apjs, 181, 244
%\bibitem[Li et al.(2019)]{Li2019} Li, Y.-P., Li, H., Ricci, L., et al.\ 2019, arXiv e-prints , arXiv:1905.01285.

\bibitem[Li et al.(2019)]{Li2019} Li, Y.-P., Li, H., Ricci, L., et al.\ 2019, \apj, 878, 39

%\bibitem[Liu et al.(2018)]{Liu2018} Liu, S.-F., Jin, S., Li, S., Isella, A., \& Li, H.\ 2018, \apj, 857, 87

\bibitem[Lidov(1962)]{Lidov1962} Lidov, M.~L.\ 1962, \planss, 9, 719.


\bibitem[Long et al.(2018)]{Long2018} Long, F., Pinilla, P., Herczeg, G.~J., et al.\ 2018, \apj, 869, 17

\bibitem[Loomis et al.(2017)]{Loomis2017} Loomis, R.~A., {\"O}berg, K.~I., Andrews, S.~M., \& MacGregor, M.~A.\ 2017, \apj, 840, 23

\bibitem[Miranda et al.(2017)]{Miranda2017} Miranda, R., Li, H., Li, S. et al. \ 2017, \apj, 835, 118

\bibitem[Miranda \& Rafikov(2019)]{Miranda2019} Miranda, R., \& Rafikov, R.~R.\ 2019, \apjl, 878, L9

\bibitem[Marzari et al.(2010)]{Marzari2010} Marzari, F., Baruteau, C., \& Scholl, H.\ 2010, \aap, 514, L4.

\bibitem[Moeckel \& Armitage(2012)]{Moeckel2012} Moeckel, N., \& Armitage, P.~J.\ 2012, \mnras, 419, 366.

\bibitem[Muto et al.(2011)]{Muto2011} Muto, T., Takeuchi, T., \& Ida, S.\ 2011, \apj, 737, 37

\bibitem[Okuzumi et al.(2016)]{Okuzumi2016} Okuzumi, S., Momose, M., Sirono, S.-. iti ., et al.\ 2016, \apj, 821, 82.

\bibitem[Papaloizou \& Larwood(2000)]{Papaloizou2000} Papaloizou, J.~C.~B., \& Larwood, J.~D.\ 2000, \mnras, 315, 823

\bibitem[Papaloizou et al.(2001)]{Papaloizou2001} Papaloizou, J.~C.~B., Nelson, R.~P., \& Masset, F.\ 2001, \aap, 366, 263

\bibitem[Pinilla et al.(2012)]{Pinilla2012} Pinilla, P., Birnstiel, T., Ricci, L., et al.\ 2012, \aap, 538, A114

\bibitem[Pinilla et al.(2015)]{Pinilla2015} Pinilla, P., de Juan Ovelar, M., Ataiee, S., et al.\ 2015, \aap, 573, A9.

\bibitem[Rafikov(2002)]{Rafikov2002} Rafikov, R.~R.\ 2002, \apj, 572, 566

\bibitem[Rasio, \& Ford(1996)]{Rasio1996} Rasio, F.~A., \& Ford, E.~B.\ 1996, Science, 274, 954.

\bibitem[Ricci et al.(2010a)]{Ricci2010a} Ricci, L., Testi, L., Natta, A., et al.\ 2010a, \aap, 512, A15

\bibitem[Rosotti et al.(2017)]{Rosotti2017} Rosotti, G.~P., Booth, R.~A., Clarke, C.~J., et al.\ 2017, \mnras, 464, L114.

\bibitem[Sefilian, \& Touma(2019)]{Sefilian2019} Sefilian, A.~A., \& Touma, J.~R.\ 2019, \aj, 157, 59

\bibitem[Shakura \& Sunyaev(1973)]{Shakura1973} Shakura, N.~I., \& Sunyaev, R.~A.\ 1973, \aap, 24, 337

\bibitem[Sheehan \& Eisner(2018)]{Sheehan2018} Sheehan, P.~D., \& Eisner, J.~A.\ 2018, \apj, 857, 18

\bibitem[Suriano et al.(2017)]{Suriano2017} Suriano, S.~S., Li, Z.-Y., Krasnopolsky, R., et al.\ 2017, \mnras, 468, 3850.

\bibitem[Suriano et al.(2018)]{Suriano2018} Suriano, S.~S., Li, Z.-Y., Krasnopolsky, R., et al.\ 2018, \mnras, 477, 1239.

\bibitem[Takahashi, \& Inutsuka(2014)]{Takahashi2014} Takahashi, S.~Z., \& Inutsuka, S.-. ichiro .\ 2014, \apj, 794, 55.

\bibitem[Takeda, \& Rasio(2005)]{Takeda2005} Takeda, G., \& Rasio, F.~A.\ 2005, \apj, 627, 1001.

\bibitem[Takeuchi \& Lin(2002)]{Takeuchi2002} Takeuchi, T., \& Lin, D.~N.~C.\ 2002, \apj, 581, 1344

\bibitem[Tanaka et al.(2002)]{Tanaka2002} Tanaka, H., Takeuchi, T., \& Ward, W.~R.\ 2002, \apj, 565, 1257

\bibitem[Tanaka \& Ward(2004)]{Tanaka2004} Tanaka, H., \& Ward, W.~R.\ 2004, \apj, 602, 388

\bibitem[Tsukagoshi et al.(2016)]{Tsukagoshi2016} Tsukagoshi, T., Nomura, H., Muto, T., et al.\ 2016, \apjl, 829, L35

%\bibitem[Teague et al.(2018)]{Teague2018} Teague, R., Bae, J., Bergin, E.~A., Birnstiel, T., \& Foreman-Mackey, D.\ 2018, \apjl, 860, L12


%\bibitem[Umebayashi \& Nakano(1981)]{Umebayashi1981} Umebayashi, T., \& Nakano, T.\ 1981, \pasj, 33, 617
\bibitem[van der Marel et al.(2019)]{vanderMarel2019} van der Marel, N., Dong, R., di Francesco, J., Williams, J.~P., \& Tobin, J.\ 2019, \apj, 872, 112

\bibitem[van der Walt et al.(2011)]{vanderWalt2011} van der Walt, S., Colbert, S.~C., \& Varoquaux, G.\ 2011, Computing in Science and Engineering, 13, 22


\bibitem[van Terwisga et al.(2018)]{Terwisga2018} van Terwisga, S.~E., van Dishoeck, E.~F., Ansdell, M., et al.\ 2018, \aap, 616, A88.

\bibitem[Wu, \& Lithwick(2011)]{Wu2011} Wu, Y., \& Lithwick, Y.\ 2011, \apj, 735, 109.

%\bibitem[Weidenschilling(1977)]{Weidenschilling1977} Weidenschilling, S.~J.\ 1977, \mnras, 180, 57

\bibitem[Yu et al.(2010)]{Yu2010} Yu, C., Li, H., Li, S., Lubow, S.~H., \& Lin, D.~N.~C.\ 2010, \apj, 712, 198

\bibitem[Zhang et al.(2015)]{Zhang2015} Zhang, K., Blake, G.~A., \& Bergin, E.~A.\ 2015, \apj, 806, L7

\bibitem[Zhu(2019)]{Zhu2019} Zhu, Z.\ 2019, \mnras, 483, 4221


\end{thebibliography}
\end{document}